\documentclass[11pt,a4paper]{article}
\usepackage{jheppub,amsmath,  amssymb,slashed,url,bm,textgreek,upgreek,hyperref}
\usepackage{graphicx}
\usepackage{epstopdf}
\def\t{{ \sf t}}

\def\tt{{\mathfrak t}}

\def\a{{\sf a}}
\def\out{{\mathrm{out}}}
\def\b{{\sf b}}

\def\HP{{H\negthinspace P}}

\def\i{{\mathrm i}}

\def\u{{\eurm u}}

\def\g{{\mathfrak g}}

\def\be{\begin{equation}}
\def\ee{\end{equation}}

\def\Im{{\mathrm{Im}}}

\def\tilde{\widetilde}

\def\h{\widehat}

\def\V{{\mathcal V}}
\def\O{{\mathcal O}}

\def\A{{\mathcal A}}
\def\d{{\mathrm d}}

\def\R{{\mathbb R}}
\def\C{{\mathbb C}}
\def\U{{ U}}

\def\[{\bigl [}

\def\]{\bigr ]}
\def\S{{\sf S}}

\def\N{{\mathcal N}}

\def\Z{{\mathbb Z}}

\def\t{\widetilde }
\def\h{\widehat}

\def\V{{\mathcal V}}

\def\M{{\mathcal M}}

\def\W{{\mathcal W}}

\def\la{\langle}

\def\Spin{{\mathrm{Spin}}}
\def\SU{{\mathrm{SU}}}

\def\ra{\rangle}

\def\H{{\mathcal H}}

\def\tilde{\widetilde}

\def\Tr{\mathrm{Tr}}
\def\TFD{{\mathrm{TFD}}}

\font\teneurm=eurm10 \font\seveneurm=eurm7  \font\fiveeurm=eurm5
\newfam\eurmfam
\textfont\eurmfam=\teneurm \scriptfont\eurmfam=\seveneurm
\scriptscriptfont\eurmfam=\fiveeurm
\def\eurm#1{{\fam\eurmfam\relax#1}}
\font\teneusm=eusm10 \font\seveneusm=eusm7 \font\fiveeusm=eusm5
\newfam\eusmfam
\textfont\eusmfam=\teneusm \scriptfont\eusmfam=\seveneusm
\scriptscriptfont\eusmfam=\fiveeusm

\font\tencmmib=cmmib10 \skewchar\tencmmib='177
\font\sevencmmib=cmmib7 \skewchar\sevencmmib='177
\font\fivecmmib=cmmib5 \skewchar\fivecmmib='177
\newfam\cmmibfam
\textfont\cmmibfam=\tencmmib \scriptfont\cmmibfam=\sevencmmib
\scriptscriptfont\cmmibfam=\fivecmmib

\def\Rr{{\sf R}}

\title{Gravity and the Crossed Product}

 \author{Edward Witten}
\affiliation{School of Natural Sciences, Institute for Advanced Study,\\ 1 Einstein Drive, Princeton, NJ 08540 USA}
\abstract{Recently Leutheusser and Liu \cite{LL,LL2} identified an emergent algebra of Type III$_1$ in the operator algebra of $\N=4$
super Yang-Mills theory for large $N$.   Here we describe some $1/N$ corrections to this picture and show that the emergent Type III$_1$ algebra becomes
an algebra of Type II$_\infty$.   The Type II$_\infty$ algebra is the crossed product of the Type III$_1$ algebra by its modular automorphism group.
In the context of the emergent Type II$_\infty$ algebra, the entropy of a black hole state is well-defined up to an additive constant, independent of the state.
This is somewhat analogous to entropy in classical physics.
}

\begin{document}\maketitle

\section{Introduction}\label{intro} 

Recently  Leutheusser and Liu \cite{LL,LL2}  studied the operator algebra of $\N=4$ super Yang-Mills theory in a novel way, arguing
that in the strict large $N$ limit, at a temperature above the Hawking-Page transition, there is an emergent operator algebra that is a von Neumann algebra of Type III$_1$.  
In the context of the thermofield double state, which is dual to a two-sided eternal black hole \cite{Malda}, the emergent Type III$_1$ algebra
and its restrictions to suitable regions of the boundary (or bulk) led to an interesting new perspective on black hole physics.

In the present article, certain $1/N^2$ corrections to this picture will be analyzed.   In the bulk dual of $\N=4$ super Yang-Mills theory, Newton's constant $G$
is proportional to $1/N^2$, so including corrections of order $1/N^2$ in the boundary theory is equivalent to including corrections of order $G$ in the bulk.
In the strict large $N$ limit, the algebra of observables on the right or left side of the black hole horizon contains a central generator, related to the black hole mass or
horizon area.   With $\O(G)$ corrections included, that is no longer the case and the right and left algebras become 
 ``factors'' in von Neumann algebra language -- that is, algebras with trivial center, analogous to a simple Lie group.
More specifically, the correction of order $G$ or $1/N^2$ that we analyze deforms the Type III$_1$ algebra of the large $N$ limit into a factor of Type II$_\infty$.  

The fact that the limiting large $N$ algebra has a nontrivial central generator is the reason that a correction perturbative in $1/N$ can qualitatively change  the algebra.
There are certainly additional contributions in the $1/N$ expansion that we do not analyze in this article, but once the center
has been eliminated, we do not expect further perturbative corrections to change the nature of the algebra.  Nonperturbative corrections are another story,
of course, since if $N$ is set to a definite integer, the algebra should be of Type I.

Mathematically, the Type II$_\infty$ algebra arises as the ``crossed product'' of the Type III$_1$ algebra of the  strict large $N$ limit by its group of modular automorphisms.
The crossed product construction has the surprising property of not depending on the cyclic separating state that is used to define it.
This fact has been important in the mathematical theory of Type III$_1$ algebras
\cite{Takesaki,Connes,ConnesBook} and is interesting physically because it might be a step toward understanding the algebra of observables outside a black
hole horizon in a ``background-independent'' way.

Deforming the algebra of operators exterior to a black hole from Type III to Type II  brings black hole physics closer to the standard framework of quantum mechanics.
The operator algebra of an ordinary quantum system is of Type I.  Type I algebras have pure states, as well as other familiar quantum concepts such as density
matrices and von Neumann entropies.   If in a bulk language,
one could describe the algebra of observables exterior to a black hole as an algebra of Type I, the black hole information
problem would very likely be solved.   The algebra of observables of a local region in quantum field theory is of Type III.   A Type III algebra does not
have pure states,  density matrices, or von Neumann entropies.   The fact that the algebra of observables outside the horizon of a black hole is of Type III in the limit
$G\to 0$
is one manifestation of the black hole information problem.   Type II algebras are intermediate between the two cases.   A Type II algebra does not have pure states,
but it does have density matrices and von Neumann entropies.    Describing the operators outside the black hole horizon by an algebra of Type II means that we
do not understand the microstates of a black hole, but we do have a framework to analyze the black hole entropy (up to an overall additive constant, as will be explained).

In section \ref{review} of this article, we will review aspects of the work of Liu and Leutheusser \cite{LL,LL2}, explaining some things in a way that will be helpful later.
In section \ref{crossed}, we introduce the crossed product construction and the deformation of the emergent Type III$_1$ algebra of the strict large $N$ limit
to an algebra of Type II$_\infty$.   We explain how to define a trace and a concept of entropy in the crossed product algebra.

As already explained, the  deformation to a Type II algebra results from properly taking into account the black hole mass or area, which one
can think of as a collective coordinate related to one of the conserved quantities that the black hole can carry, namely its energy.    The black hole can carry
other conserved charges, notably angular momentum and gauge charges, and a fuller description must include additional collective coordinates.
 As explained in section \ref{symmetries}, this requires making a crossed
product not just with the modular automorphism group but with a larger group of automorphisms.  This more extended version of the crossed product, however,
does not lead to a further qualitative change in the algebraic description.

Some background that might be helpful in understanding the present article can be found in \cite{Lecture,OlderLectures}.

\section{The Large $N$ Limit}\label{review}

With a convenient normalization of the fields, the action of $\N=4$ super Yang-Mills theory with gauge group $SU(N)$ is conveniently written
$N\Tr\, L$, where $L$ is a gauge-invariant polynomial in the fields and their derivatives (with no explicit factors of $N$).   With this normalization, correlation
functions of any single-trace operator $\W=\Tr\,W$, where $W$ is a gauge-invariant polynomial in the fields and their derivatives (again with no explicit
powers of $N$), have a simple large $N$ scaling.
In the vacuum or in a thermal state, $\W$ has an expectation value of order $N$, a connected two-point function of order 1, and more generally
a connected $k$-point function of order $N^{2-k}$.   Therefore, to define a large $N$ limit of the operator algebra, as analyzed by Leutheusser and Liu
(LL) in \cite{LL,LL2},  we consider subtracted single-trace operators of the form $\W=\Tr\,W -\la \Tr\,W\ra$, with all possible choices of $W$.   Since $\la\Tr\, W\ra$ 
diverges for $N\to\infty$, the subtraction is necessary in order to get a large $N$ limit.

 \begin{figure}
 \begin{center}
   \includegraphics[width=3.5in]{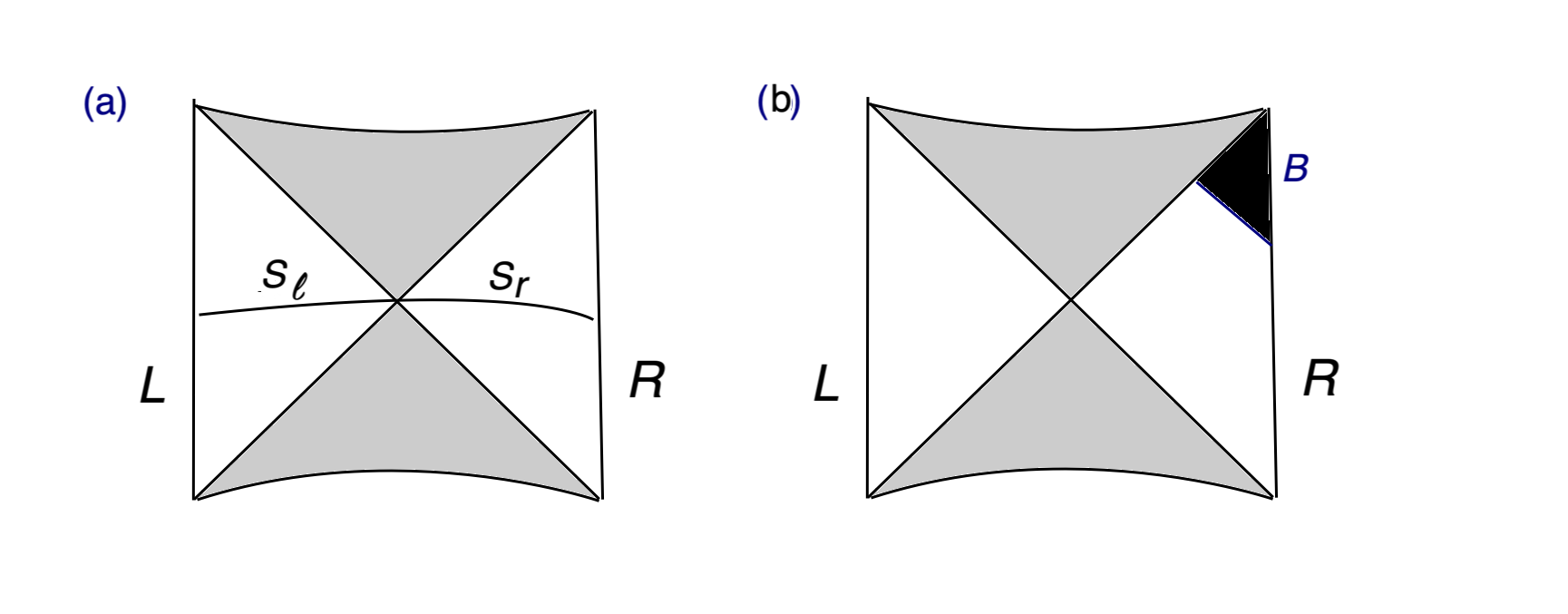}
 \end{center}
\caption{\small  (a)  A two-sided eternal black hole, with shaded regions to the future of the future horizon and the past of the past horizon.
 Operators on the right and left boundaries, labeled $R$ and $L$, describe physics
in the unshaded regions to the right or left of the black hole, respectively.   Also sketched is a Cauchy hypersurface $S$ that passes through
the bifurcate horizon.   The horizon divides $S$ in a natural way into left and right portions $S_\ell$ and $S_r$. (b) A time band $B=(t_0,\infty)$ of the right
boundary.   According to \cite{LL,LL2}, in the large $N$ limit, operators in $B$ describe physics in the causal wedge
of $B$, which is sketched in black.  \label{BH}}
\end{figure} 

 For example, LL consider a thermal state at temperature $T$, on a spatial manifold $\S^3$.   They are primarily
interested in the case that $T$ exceeds the Hawking-Page temperature
$T_\HP$, or equivalently  that $\beta=1/T$ is less than $\beta_\HP=1/T_\HP$,
because in this case  a thermal ensemble on the boundary of AdS$_5$ is dual to a black hole in the interior.    In fact, they consider a pair of
copies of the theory, the ``right'' and ``left'' copies R and L, entangled in the thermofield double state $\Psi_\TFD$ at temperature $T>T_\HP$, and dual to
a two-sided eternal black hole \cite{Malda}; see fig. \ref{BH}(a).
For $T>T_\HP$, the free energy of $\N=4$ super Yang-Mills theory is of order $N^2$ for large $N$, with a $T$-dependent coefficient,
just as in the case of QCD \cite{Thorn}.
  Likewise the terms of order $N$ in the thermal expectation value $\la\Tr\,W\ra$ (for various choices of $W$) are $T$-dependent in the high temperature phase, and
  therefore the algebra defined by LL depends explicitly on $T$.
Similarly, if we introduce chemical potentials for other conserved charges that the black hole can carry -- angular momentum and the charges associated
with the $\Rr$-symmetry group $SU(4)_\Rr$ of $\N=4$ super Yang-Mills --  the large $N$ limit of the algebra of single-trace operators depends on those chemical 
potentials.

As already remarked, the subtracted single-trace operators have Gaussian correlation functions:  two-point functions  of order 1, and $k$-point functions
for $k>2$ that vanish at $N=\infty$.   Therefore, in the large $N$ limit,  commutators of single-trace operators (and their anticommutators, in the case of
fermionic operators) are $c$-numbers.   The matrix
of commutators has a finite-dimensional kernel, as explained presently.   For the moment we postpone discussion of the kernel and consider only the single-trace
operators that have nonzero commutators.     Because those commutators are $c$-numbers, the subtracted single-trace operators describe a generalized free field theory
in the large $N$ limit,  as explained by LL.
From bounded functions of the operators of the generalized free field theory, we can make a von Neumann algebra $\A$. This is the emergent  von Neumann
algebra defined in \cite{LL,LL2}.

In the thermofield double state, we really have two copies of this algebra, the algebras $\A_{R,0}$ and $\A_{L,0}$ of the right and left system.   (The subscript
0 means that we do not yet include the operators with trivial commutators; it will be removed when we include them.)
LL give two explanations of the fact that $\A_{R,0}$ and $\A_{L,0}$ are von Neumann algebras of\footnote{An introduction to the different
types of von Neumann algebra can be found, for example, in section 3 of \cite{Lecture} or section 6 of \cite{OlderLectures}.}  Type III$_1$.   One argument uses
AdS/CFT duality.   $\A_{R,0}$ and $\A_{L,0}$ are dual to the algebras $\A_{r,0}$ and $\A_{\ell,0}$ of low energy effective field theory outside the black hole
horizon, on the right and left side respectively:
\be\label{firstside} \A_{R,0}=\A_{r,0},~~~ \A_{L,0}=\A_{\ell,0}. \ee    
It will become clear that not including the central generators in $\A_{R,0}$ and $\A_{L,0}$ is dual to defining $\A_{r,0}$ and $\A_{\ell,0}$ in terms of
low energy effective field theory in the field of a black hole of definite mass.
   When we extend $\A_{r,0}$ and $\A_{\ell,0}$ to  let the mass vary, we will remove the subscript 0.
Like the algebra of any local region  in quantum field theory, $\A_{r,0} $ and $\A_{\ell,0}$ are of Type III$_1$.   See footnote \ref{noted} below for
a second explanation that the algebras are of Type III$_1$. 

A black hole carries conserved charges -- it has a mass or energy, and it can carry angular momentum or conserved gauge charges.\footnote{In the context
of $\N=4$ super Yang-Mills theory, the conserved gauge charges are dual to the $\Rr$-symmetry charges of
the boundary theory.}    These charges are determined by the temperature and chemical potentials of the boundary theory.  Since $\A_{L,0}$ and $\A_{R,0}$ depend on
the temperature and the chemical potentials, likewise $\A_{\ell,0}$ and $\A_{r,0}$ depend on the black hole mass and other conserved charges.
 The mass and other conserved charges carried by 
 the black hole are  parameters of the classical black hole solution.  One can think of those parameters as collective coordinates, which are shared between the left and right of the black hole and are 
not part of what is described by the left or right bulk algebras $\A_{\ell,0}$ and $\A_{r,0}$.   $\A_{\ell,0}$ and $\A_{r,0}$  describe small
fluctuations to the left or right of the black hole horizon, for particular values of the collective coordinates.   

Of the conserved charges carried by the black hole, the one that is really important for a qualitative understanding is the mass or energy.   We will postpone
discussing the other conserved charges until section \ref{symmetries}.

Since the black hole mass is not a generator of the bulk algebra $\A_{\ell,0}$ or $\A_{r,0}$, the gauge theory Hamiltonians $H_L$ or $H_R$, which generate
translations of the time coordinate $t$ along the left or right boundaries, must not be part of the boundary algebra
$\A_{L,0}$ or $\A_{R,0}$.   The mechanism by which this comes about is straightforward.  The Hamiltonian of $\N=4$ super Yang-Mills theory, like the Lagrangian,
is defined with an explicit factor of $N$:
\be\label{toldo} H_R=\frac{N}{2 g^2} \int_\S \d^3x \sqrt g \Tr\, \left( F_{0i} F_{0j} g^{ij}+\cdots \right), \ee
where various additional terms are omitted.  Above the Hawking-Page transition, because of the explicit factor of $N$,  $H_R$ has a thermal expectation value 
of order $N^2$ and also a connected two-point function of order $N^2$.   We can of course subtract the expectation value and consider the operator
$H'_R=H_R-\la H_R\ra$.   But because $\la H_R'{}^2\ra\sim N^2$, $H_R'$ does not have a large $N$ limit.\footnote{This is analogous
to the fact that in quantum statistical mechanics, the usual Hilbert space and Hamiltonian do not have an infinite volume limit (even if one subtracts a
constant from the Hamiltonian so that its thermal expectation value vanishes). This is why the thermofield double
was introduced \cite{HHW}.  It provides a way to describe the infinite volume limit in a Hilbert space.   See for example  \cite{Lecture}.}

If  we simply divide $H_R'$ by $N$, we get an operator with bounded fluctuations that
does have a large $N$ limit: 
\be\label{fugo} \U=\frac{1}{N}H_R'. \ee
The correlation functions of $\U$ have the same large $N$ scaling as any other single-trace operator that is defined with no explicit power of $N$,
so in particular $\U$ has a large $N$ limit.   However, in the large $N$ limit, $\U$ is central.   That is because, for any $\V\in \A_R$,
\be\label{nugo} [\U,\V]=\frac{1}{N}[H_R,\V]=-\frac{\i}{N}\frac{\partial \V}{\partial t}. \ee
Thus, at $N=\infty$, $\U$ becomes central and commutes $\A_{R,0}$.  Since we defined $\A_{R,0}$ to consist only of the single-trace operators that have nontrivial
commutators,   $\U$ is not part of $\A_{R,0}$.   Shortly, we will   define an extended algebra $\A_R$ with $\U$ as an additional (central) generator.

Because $H_R$ is not part of $\A_{R,0}$, time translations are a group of outer automorphisms\footnote{Similarly, in the large volume limit of quantum
statistical mechanics, time translations are a group of outer automorphisms of the algebra of observables that acts on the thermofield double Hilbert space
\cite{HHW}; see \cite{Lecture} for an explanation.}
of that algebra.  This fact is very important in the work of LL.  Let 
$B$ be a ``time band'' $t_0<t<t_1$ in the right boundary, 
with $t_0,t_1\in \R\cup\pm\infty$.   Let $\A_{R,0}^B$ be the algebra generated by subtracted single-trace operators in $B$.
Any local operator can be conjugated into the time band by a time translation.   So if time translations were part of $\A_{R,0}$,
$\A_{R,0}^B$ would simply coincide with $\A_{R,0}$.  Since the generator of time translations is actually not part of $\A_{R,0}$, $\A_{R,0}^B$ is a proper subalgebra of $\A_{R,0}$.
 As explained by LL, the algebra $\A_{R,0}^B$ is dual to the bulk algebra of operators in the ``causal wedge'' of
the time band $B$ (fig. \ref{BH}(b)), and thus is of Type III$_1$, like the bulk algebra of any local region. 
  If one backs away from the strict large $N$ limit and considers an expansion in powers of $1/N$, then as asserted in \cite{LL,LL2}, the definition
of $\A_{R,0}^B$ makes sense to all finite orders in $1/N$, since if $\phi(\vec x,t)$ is a local operator in a given time band, its time derivatives of any finite order
are also local operators in the same time band.    

Although the operators $H_R$ and $H_L$ do not have large $N$ limits, the difference $\h H=H_R-H_L$ does have such a limit.   Indeed, $\h H$ annihilates
the thermofield double state $\Psi_\TFD$, so in particular it does not have  divergent fluctuations in that state, in contrast to $H_R$ and $H_L$.   From a boundary point of view, the thermofield double Hilbert space $\H_\TFD$ is defined as the Hilbert space spanned by  the states  $\a\Psi_\TFD$ with
$\a\in \A_{R,0}$. (It can equally well be defined as the Hilbert space spanned  by states $\a\Psi_\TFD$ with $\a\in\A_{L,0}$; this gives the same Hilbert space.) 
From a bulk point of view, $\H_\TFD$ is more simply the Hilbert space obtained by quantizing the small fluctuations about the eternal black hole
(for fixed values of the collective coordinates).
The fact that $\h H=H_R'-H_L'$ exists as an operator acting on $\H_\TFD$ but $H_R'$ and $H_L'$ do not shows that $\H_\TFD$ is not a simple tensor 
product\footnote{This fact has an analog in
the thermodynamic limit of quantum statistical mechanics \cite{HHW}; see \cite{Lecture}.} of left and right Hilbert spaces acted on by $\A_{L,0}$ and $\A_{R,0}$.  This nonfactorization
of $\H_\TFD$ in the large $N$ limit is also related to the  fact that $\A_{L,0}$ and $\A_{R,0}$, defined in the large $N$ limit, are not of Type I.   If these algebras were of
Type I, they would have irreducible representations $\H_R$ and $\H_L$, and we  could hope for a factorization $\H_\TFD=\H_L\otimes \H_R$, similarly
to what actually happens for large $N$ below the Hawking-Page temperature.   As the algebras are of Type III, there are no candidates for $\H_L$ or $\H_R$.

What is the bulk dual of the fact that $H_R'=H_R-\la H_R\ra$ and $H_L'=H_L-\la H_L\ra$ do not have large $N$ limits, but $\h H=H'_R-H'_L$ does
have such a limit?   The eternal black hole has a Killing vector field $V$ that generates time translations.  $V$  is future-directed timelike 
on the right side of the black hole outside the horizon
and past-directed timelike on the left side of the black hole outside the horizon.   The associated conserved charge is
\be\label{tolgo}\h h=\int_S \d\Sigma^\mu V^\mu T_{\mu\nu}, \ee
where $S$ is a Cauchy hypersurface of the bulk theory, and $T_{\mu\nu}$ is the energy-momentum tensor of the bulk fields (including the energy-momentum
pseudo-tensor of the bulk gravitational fluctuations).   The operator $\h h$ has a simple boundary dual
\be\label{zolgoox}\h h=\beta \h H. \ee
 Eqn. (\ref{zolgoox}) reflects the fact that the vector field $V$ reduces to $\beta\partial_t$
 on the right boundary, and to $-\beta\partial_t$ on the left boundary.
To try to split $\h h$ as the difference of a ``right'' part and a ``left'' part, one can proceed as follows.   Choose the Cauchy hypersurface $S$ to pass through
the bifurcate horizon where the left and right exteriors of the eternal black hole meet (fig \ref{BH}(a)), and write $S=S_\ell\cup S_r$, where $S_\ell $ and $S_r$ are the left
and right portions of $S$.   Then define
\begin{align}\label{nolgo} h_r& =\int_{ S_r}\d\Sigma^\mu V^\nu T_{\mu\nu}\cr  
                                          h_\ell& =-\int_{ S_\ell}\d\Sigma^\mu V^\nu T_{\mu\nu}\cr  
                                          \h h&=h_r-h_\ell.\end{align}
 (A minus sign is included in the definition of $h_\ell$ because $V$ is past-directed in the left region; with this minus sign included, $h_\ell$ formally propagates
 states towards the future.)    These expressions formally split $\h h$ as a difference between right and left operators.  The problem with the splitting is that, because of
 divergent fluctuations,  $h_r$ and $h_\ell$
 do not make sense as operators.      Actually,  $h_r$ and $h_\ell$ do make sense as quadratic forms; that is, for suitable states $\Psi,\Psi'\in\H_\TFD$, the
 matrix elements $\la\Psi'|h_r|\Psi\ra$ and $\la\Psi'|h_\ell|\Psi\ra$ are well-defined.   But $h_r$ and $h_\ell$ do not make sense as operators on 
 $\H_\TFD$, because for example, due to an ultraviolet divergence near the horizon, $|h_r|\Psi_\TFD\ra|^2=\la\Psi_\TFD|h_r^2|\Psi_\TFD\ra=\infty$, so $h_R|\Psi_\TFD\ra$
 is not a Hilbert space state.   More generally, for any $\Psi\in\H_\TFD$, $h_r\Psi$ is not square-integrable and so not in $\H_\TFD$.
 All this is in parallel with properties of the boundary operators $H_R'$ and $H_L'$, which likewise make sense as quadratic forms but, because of fluctuations, do
 not make sense as operators on $\H_\TFD$.

The modular operator\footnote{See for example \cite{OlderLectures} for basic facts about modular operators.} of the state $\Psi_\TFD$ for the algebra $\A_{R,0}$ is $\Delta=e^{-\beta\h H}$.   (Exchanging $\A_{R,0}$ and $\A_{L,0}$ reverses the sign of $\h H$, so the
modular operator of $\A_{L ,0}$ for the state $\Psi_\TFD$ is $\Delta^{-1}=\exp(+\beta\h H)$.)  
 One way to deduce the formula $\Delta=e^{-\beta\h H}$ is to start with finite $N$, where  the Hilbert space
 of the thermofield double is a simple tensor product $\H_L\otimes \H_R$, and the 
algebras of observables on the left and right factors
are of Type I.   In this description, the left and right Hamiltonians are $H_L=H\otimes 1$, $H_R=1\otimes H$, where $H$ is the usual Hamiltonian on a single
copy of the system.   In such a Type I situation, a state $\Psi\in \H_L\otimes \H_R$ has reduced density matrices $\rho_R$ and $\rho_L$ for the right and left systems
and the modular operator of $\Psi$ for $\A_R$ is a tensor product $\Delta=\rho_L^{-1}\otimes \rho_R$.    In the case $\Psi=\Psi_\TFD$, the density matrices are thermal:
$\rho_R=\rho_L=e^{-\beta H}/Z$, where $Z$ is the partition function, so $\Delta=\exp(-\beta H_R+\beta H_L)=\exp(-\beta \h H)$.   Both $\Delta $ and $\h H$ have a large $N$
limit, and,  since  the relation  $\Delta=e^{-\beta \h H}$ holds for every $N$, it automatically is valid in the large $N$ limit.   By contrast, the intermediate steps in the derivation involved operators  $\rho_R $
and $\rho_L$ and Hilbert spaces $\H_R$ and $\H_L$ that do not have large $N$ limits.  
 The formula $\Delta=e^{-\beta \h H}$ with $\h H\Psi_\TFD=0$ implies
\be\label{ingo} \Delta \Psi_\TFD=\Psi_\TFD.\ee
Since $\beta \h H$ is equal to the bulk operator $\h h$, the bulk counterpart of the boundary statement $\Delta=e^{-\beta \h H}$ is simply $\Delta = e^{-\h h}$.  
The result $\Delta = e^{-\h h}$ can also be deduced from a classical result of Bisognano and Wichman about Rindler space \cite{BW}, 
or more exactly from the analog of this result for the eternal black hole.\footnote{This analog was first described by Sewell \cite{Sewell},
who also observed a close relation to 
Unruh's thermal interpretation of Rindler space \cite{Unruh}.}

LL in \cite{LL,LL2} exploited the relation $\Delta=e^{-\beta\h H}$ or equivalently $\beta \h H=-\log \Delta$ in the following way.  
Conjugation by $e^{\i u\h H}$, $u\in \R$ is called modular flow.   Since $\h H$ generates time translations on the right boundary, 
modular flow for the algebra $\A_{R,0}$ with parameter $u$ shifts the boundary time by $t\to t+\beta u$.     Let $B$ be the time band $t_0<t <\infty$ (fig. \ref{BH}(b)).   Modular flow for
the algebra $\A_{R,0}$ with flow parameter $u>0$ maps $B$ to itself, and therefore it conjugates $\A_{R,0}^B $ to itself.   Such a ``half-sided modular inclusion'' of von Neumann algebras\footnote{\label{noted}Half-sided modular inclusions exist only for algebras of Type III$_1$ \cite{Borch}.   As observed in \cite{LL,LL2}, the existence of the half-sided modular inclusion $\A_{R,0}^B\subset \A_{R,0}$  gives an explanation of the Type III$_1$ nature
of $\A_{R,0}$  that does not require duality with a bulk description.}   has very strong
implications, which LL use in proposing a way to probe behind the black hole horizon.     One notable fact about their construction is the following.    As LL explain,
the half-sided modular inclusion in this problem can be seen to lowest order in $G$ directly in the bulk description.   Purely from a bulk point of view,
it would not be clear that this structure persists beyond lowest order, when quantum fluctuations of the spacetime are taken into account.    However,
in the boundary theory, it is obvious that the definitions of $\A_{R,0}$ and $\A_{R,0}^B$ and the statements about a half-sided modular inclusion make sense
to all orders in $1/N$.  Hence from a bulk point of view, this structure exists to all orders in $G$.   

The state $\Psi_\TFD$ for the algebra $\A_{R,0}$ is defined so that thermal expectation values are the same as expectation values in the state $\Psi_\TFD$:
\be\label{wonko} \la \a\ra=\la\Psi_\TFD|\a|\Psi_\TFD\ra,~~~ \a\in \A_{R,0}.\ee
The same formula holds for $\a\in\A_{L,0}$.   Defined this way, $\Psi_\TFD$ describes expectation values of operators in $\A_{R,0}$ and $\A_{L,0}$, or their
bulk duals $\A_{r,0}$ and $\A_{\ell,0}$, without taking into account the central operator $\U$.   Let us define an extended algebra $\A_R$ that includes
$\U$.   
We define $\A_R$ as a tensor product 
\be\label{deften}\A_R=\A_{R,0}\otimes \A_\U,\ee
 where $\A_\U$ is the abelian algebra of bounded functions of $\U$.
While $\A_{R,0}$ acts on  $\H_\TFD$, $\A_R$ acts on $\h\H_\TFD=\H_\TFD\otimes L^2(\R)$, where $L^2(\R)$ is the space of square-integrable
functions of $\U$, with $\U$ acting by multiplication. 
An element of $\A_R$ is a $\U$-dependent element of $\A_{R,0}$ such as
\be\label{ygo}\h \a =\int_{-\infty}^\infty \d u \,\a(u) e^{\i u \U},~~~~ \a(u)\in\A_{R,0}\ee
(here $\a(u)$ need not be a smooth function of $u$; for example, we may have $\a(u)=\a \delta(u-u_0)$ for some $\a, u_0$, corresponding to
$\h\a=\a e^{\i u_0\U}$). 
We can define an extended thermofield double state that describes thermal expectation values of elements of the extended
algebra $ \A_R$.   To do this, we simply note that, as its connected $k$-point functions vanish in the large $N$ limit for $k>2$, the thermal correlation
functions of $\U$ are Gaussian.   In other words, there is a Gaussian function $g(\U)$ such that for any $k$, $\la \U^k\ra=\int_{-\infty}^\infty \d \U \,g(\U) \U^k$.
By definition, $\U=(H_R-\la H_R\ra)/N$ has mean 0; its variance is the heat capacity $\la \U^2\ra=-(1/N^2)\partial_\beta \la H_R\ra.$  So the Gaussian
function $g(\U)$ is concretely
\be\label{gaussf} g(\U)=\left(\frac{N^2}{2\pi |\partial_\beta\la H_R\ra |}\right)^{1/2} \exp\left(-\frac{N^2 \U^2}{2|\partial_\beta \la H_R\ra|}\right). \ee
We can define an extended thermofield double state $\h\Psi_\TFD= g(\U)^{1/2}\Psi_\TFD$ that captures thermal expectation values of operators in the
extended algebra $\A_R$, in the sense that for $\h \a$ as in (\ref{ygo}), we have
\be\label{wygo}\la\h \a\ra=\la\h\Psi_\TFD|\h \a|\h\Psi_\TFD\ra=\int_{-\infty}^\infty \d \U\,g(\U) \int_{-\infty}^\infty\d u\,e^{\i u\U} \la\Psi_\TFD|\a(u)|\Psi_\TFD\ra.\ee
Because $\U$ is central in the algebra $\A_R$, the state $\h\Psi_\TFD$ is what is called a classical-quantum state.   In general, the definition of a classical-quantum
state is as follows.  Consider a bipartite quantum system $AB$ with Hilbert space $\H_{AB}=\H_A\otimes \H_B$.   (In our application, $\H_{AB}=\h\H_\TFD$, $\H_A=L^2(\R)$ and
$\H_B=\H_\TFD$.)  A classical-quantum  density matrix on $\H_{AB}$ is a density matrix of the form $\rho_{AB}=\sum_i p_i |i\ra_A\la i|_A\otimes \rho_{i,B}$, for some basis $|i\ra_A$ of $\H_A$,
some density matrices $\rho_{i,B}$ on $\H_B$, and some positive numbers $p_i$ with $\sum_i p_i=1$.    Equivalently, $\rho_{AB}$ commutes with
a maximal set of commuting operators of the $A$ system (the diagonal matrices in the basis $|i\ra_A$).  Because $\U$ is central, not only is $\h\Psi_\TFD$ a classical-quantum state, but actually any
state of the algebra $\A_R$ is classical-quantum.   When we deform away from the large $N$ limit in section \ref{crossed}, there will no longer be a 
 central element $\U$,  and the generic state will no longer be classical-quantum.

In this discussion, we extended the algebra $\A_{R,0}$ by adding another generator $\U=H_R'/N$.   What happens if we also define $\U'=H_L'/N$ and add it to $\A_L$? $\U'$ is not really ``new''
since $\U-\U'=\h H/N$, where $\h H$ has a large $N$ limit, so that  $\h H/N$ vanishes at $N=\infty$.
   So to get a proper arena for the action of all boundary operators in the large $N$ limit, 
we only have to add one new generator $\U$ associated to fluctuations in energy 
(plus additional ones that are related in the same way to the angular momentum and conserved $\Rr$-charges; see section \ref{symmetries}).  
Thus to the algebras $\A_{L,0}$ and $\A_{R,0}$, we only add one new generator $\U$ which is in a sense ``shared'' between the two sides, rather as from a bulk
point of view the horizon area $A$ is visible from either side of the horizon and thus is ``shared.'' 
In the large $N$ limit,
\be\label{weften}\A_L=\A_{L,0}\otimes \A_\U,\ee
in perfect parallel with eqn. (\ref{deften}).  With $1/N$ corrections included, matters are more subtle, as we will see in section \ref{crossed},
 and it is no longer true that $\A_L$ and $\A_R$ have in common an
operator such as $U$.

As explained in \cite{LL,LL2}, the algebra $\A_R$ is of Type III in the large $N$ limit.   However, it is not a ``factor'' in von Neumann algebra language
because its center does not just consist of the complex scalars $\C$.  Rather, its 
center is the infinite-dimensional commutative algebra $\A_\U$.   It may seem somewhat nongeneric that $\A_R$ has such a large center at $N=\infty$.
We will learn in section \ref{crossed}  that $1/N$ corrections modify the algebra so that its center becomes trivial.  In the process, the algebra
will be deformed from being of Type III to being of Type II.   Indeed, it will become a factor of Type II$_\infty$.

We conclude with a few facts that will be useful in section \ref{crossed}.   First, if an algebra $\A$ acts on a Hilbert space $\H$, a vector $\Psi\in\H$
is said to be cyclic for $\A$ if $\H$ is spanned by vectors $\a\Psi$, $\a\in\A$.  It is said to be separating for $\A$ if $\a\Psi=0$, $\a\in\A$, implies $\a=0$.
The vector $\h\Psi_\TFD$ is separating for $\A_R$ because, as the thermal expectation value of any operator $\h\a^\dagger\h\a$ is always strictly positive
for any $\h\a\not=0$, eqn. (\ref{wygo}) implies that $\h a\h\Psi_\TFD\not=0$ for all nonzero $\h\a\in \A_R$.   $\h\Psi_\TFD$ is also cyclic for $\A_R$ because (1)
$\H_\TFD$ was defined in a way that ensured that $\Psi_\TFD$ is cyclic for $\A_{R,0}$; (2) since the Gaussian function $g(\U)$ is everywhere positive,
any function $f(\U)$ can be factored as $m(\U) g^{1/2}(\U)$ for some function $m(\U)$, and therefore the state $g^{1/2}(\U)$ is cyclic for the abelian
algebra $\A_\U$, acting on $L^2(\R)$.   The statement that the product state $\h\Psi_\TFD $ is cyclic for $\A_R=\A_{R,0}\otimes \A_\U$ 
is just a composite
of these two statements.

\section{The $1/N$ Expansion and the Crossed Product}\label{crossed}

\subsection{Beyond the Large $N$ Limit}\label{beyond}

Our goal in this section is to understand $1/N$ corrections to the picture described in section \ref{review}. 
In section \ref{review}, we considered an algebra generated by single-trace operators with their thermal expectation values subtracted, an example
being $\W=\Tr\,W-\la\Tr\, W\ra$, where $W$ is a polynomial in the fields and their derivatives.   A typical element of the algebra was a complex linear combination of products of such operators.  To depart from the strict large $N$
limit, we have to modify the definition slightly so that coefficients are not complex numbers, but rather functions of $N$ that have an asymptotic expansion 
in powers of $1/N^2$ around the $N=\infty$ limit.   Thus an element of the algebra is a sum of elements
\be\label{donkey} \sum_{k=0}^\infty a_k N^{-2k}\W_1\W_2\cdots \W_s,~~~~~a_k\in \C. \ee   (The  coefficients $a_k$ depend on the choice of 
$\W_1,\cdots,\W_s$.   The associated von Neumann algebra is generated by bounded functions of  such expressions.)   
Another way to say this is that we work not over $\C$ but over a formal power series ring $\C[[1/N^2]]$.
This is necessary because operator product coefficients and commutation relations of $\N=4$ super Yang-Mills theory have nontrivial asymptotic expansions
 in powers of $1/N^2$.   To define an algebra that is consistent with operator product expansions and commutators, we have to allow
the operators to have  $N$-dependent coefficients of this form.

In section \ref{review}, we described the algebra of bulk operators in the field of a black hole, in the large $N$ limit, 
in terms of the left and right algebras $\A_{r,0}$ and $\A_{\ell,0}$
and a central generator $\U$.   In this description, the Hilbert space is $\h\H_\TFD=\H_\TFD\otimes L^2(\R)$, where $\A_{r,0}$ and $\A_{\ell,0}$ act on $\H_\TFD$,
and  $L^2(\R)$ is the space of square-integrable functions of $\U$, with $\U$ acting by multiplication.
We also observed that the boundary operator $H'_R/N$ makes sense in the large $N$ limit as an element of the right algebra $\A_R$,
and  in the large $N$ limit we identified $H'_R/N$ with $\U$.

Once we go beyond the $N=\infty$ limit, $H_R'/N$ is no longer central.   Rather, for any $\a=\A_R$, we have $[H'_R/N,\a]=(-\i/N)\partial_t \a$.
To identify a bulk dual of $H_R'/N$, we have to find a bulk operator that satisfies the same commutation relation.   Such an operator presents itself,
namely the operator ${\frac{1}{\beta}\h h}$, where $\h h$ was defined in eqn. (\ref{nolgo}).  This operator satisfies the desired commutation relation, because $\h h$  is the conserved
charge associated to a Killing vector field that coincides with $\beta\partial_t$ on the right boundary.   This suggests the identification
\be\label{idgo}\frac{1}{N}H'_R=\U+\frac{1}{\beta N}\h h. \ee
Formally, one might be tempted to use here $h_r/\beta$ rather than $\h h/\beta$, on the grounds that the two operators have the same commutation relations with operators
that are to the right side of the black hole.   However, as was explained earlier, $h_r$ is not a well-defined operator.    We will  get 
more insight by expressing all statements in terms of operators that actually exist.  

The left and right hand sides of eqn. (\ref{idgo}) have the desired commutation relations with elements of $\A_R$, but this does not determine the right hand side
uniquely: without spoiling anything, we could add  an arbitrary element\footnote{We cannot add an element of $\A_{r,0}$, because such an element is noncentral
and would contribute to the commutators with elements of $\A_{r,0}$.} $\b_\ell\in \A_{\ell,0}$ or an arbitrary function $f(\U)$.  
 More generally, we could add any $\h b \in \A_{\ell,0}\otimes \A_\U$.   Thus we should ask whether eqn. (\ref{idgo}) should be corrected to \be\label{nidgo}\frac{1}{N}H'_R\overset{?}{=}\U+\frac{1}{N\beta }\h h+\frac{1}{N}\h b +\O(1/N^2),\ee
where we also include the possibility of corrections in higher orders in $1/N$.   The answer is that  to reconstruct $\A_R$,  there is no need to
add any such corrections, because they can be removed by conjugation.   Thus, we introduce the operator $\Pi=-\i\d/\d\U$, acting on $L^2(\R)$.
Conjugation by $\exp(-\i\h b\Pi/N)$ will remove the $\h b/N$ term from the right hand side of eqn. (\ref{nidgo}), and order by order in $1/N$, a similar conjugation would remove
any possible higher order terms in $1/N$ that could be added in eqn. (\ref{nidgo}) without spoiling the commutators with $\A_R$.  

Since $\h h=\beta(H_R'-H_L')$, eqn. (\ref{idgo}) also implies
\be\label{widgo}\frac{1}{N}H_L'=U. \ee

Thus, order by order in $1/N$, we can interpret the boundary algebra $\A_R$ as the algebra generated by $\A_{r,0}$ together with one
more operator $\U+\h h /\beta N$, acting on $\H_\TFD\otimes L^2(\R)$.    
We will summarize this by saying that order by order in $1/N$,
$\A_R=\A_{r,0}\rtimes \A_{\U+\h h/\beta N}$.    Here $ \A_{r,0}\rtimes \A_{\U+\h h/\beta N}$ is the algebra generated by $\A_{r,0}$ together with ${\U+\h h/\beta N}$,
and the meaning of the  symbol $\rtimes$ is that $\U+\h h/\beta N$ does not commute with $\A_{r,0}$ but rather generates an outer automorphism of $\A_{r,0}$.   
 At the von Neumann algebra level, we should consider
not $\U+\h h/\beta N$ but bounded functions of this operator.  However, we do not indicate this in the notation.

This description of $\A_R$, however, is only perturbative in $1/N$.   It may not hold if we set $N$ to a definite integer value such as 101,
 because the series of conjugations that was used to set $H'_R/N=\U+\h h/\beta N$ order by order may not converge at integer $N$.  
In fact, we expect that to happen, because the description just given
 leads to a Type II algebra, as we will see, but when $N$ is an integer, $\A_R$ is of Type I.
 
 Going back to perturbation theory in $1/N$, the reader may observe an apparent difficulty.
 $\A_R$ is supposed to commute with $\A_L$, which in the large $N$ limit was identified with $\A_{\ell,0}\otimes \A_\U$.      Our proposal 
 $\A_R=\A_{r,0}\rtimes \A_{\U+\h h/\beta N}$ is not
 consistent with that identification of $\A_L$.   That is because, although $\A_{r,0}$ commutes with $\A_{\ell,0}\otimes \A_\U$, $\A_{\U+\h h/\beta N}$ does not;
 $\U+\h h/\beta N$ commutes with $\U$ but not with $\A_{\ell,0}$.   There is, however, a simple fix.   Although $\U+\h h/\beta N$ does not commute with $\a_{\ell}\in \A_{\ell,0}$,
 it does commute with $e^{\i\Pi \h h/\beta N}\a_\ell e^{-\i\Pi \h h/\beta N}$.   So we propose that $\A_L$ is generated order by order in $1/N$ by $\U$ (or more exactly,
 by bounded functions of $\U$) along with operators of the form  $e^{\i\Pi \h h/\beta N}\a_\ell e^{-\i\Pi \h h/\beta N}$.   More succinctly, we describe this by writing
 $\A_L=e^{\i\Pi \h h/\beta N}\A_{\ell,0} e^{-\i\Pi \h h/\beta N}\rtimes \A_\U$.   
 
The underlying thermofield double state is symmetrical between $\A_L$ and $\A_R$, but there is an  apparent asymmetry between these descriptions of $\A_L$ and $\A_R$.   This asymmetry, however, can be removed
 by conjugating both algebras with $e^{-\i \Pi \h h/2N}$.    This leads to  \begin{align}\label{elb}\A_R&= e^{-\i \Pi\h h/2\beta N}\A_{r,0} e^{\i \Pi\h h/2\beta N}\rtimes \A_{\U+\h h/2\beta N}\cr
 \A_L&=e^{\i\Pi\h h/2\beta N}\A_{\ell,0} e^{-\i \Pi \h h/2\beta N}\rtimes \A_{\U-\h h/2\beta N}.\end{align}   Since  $\h h$ is odd under the exchange of $\A_{\ell,0}$ and $\A_{r,0}$, these formulas
 treat $\A_L$ and $\A_R$ symmetrically.
 
 However, the symmetrical formulas are slightly less convenient, so we instead will use the description
 \begin{align}\label{nalign} \A_R& =\A_{r,0} \rtimes \A_{\U+\h h/\beta N} \cr
                      \A_L&=e^{\i\Pi \h h/\beta N} \A_{\ell,0} e^{-\i \Pi \h h/\beta N}\rtimes \A_\U, \end{align}
  valid to all orders in $1/N$. 
  
  From this description, we can see at least heuristically 
  that the $1/N$ corrections have deformed $\A_R$ so that its center has become trivial.   The central element $\U$ of $\A_R$ has been deformed to 
  $\U+\h h/\beta N$, an element of  $\A_R$ that is not central.   It is true that the deformed $\A_R$ commutes with $\U$, but $\U$ is not an element of $\A_R$,
  so it is not in the center of $\A_R$.  Rather, $\U$ is an element of $\A_L$, which explains why it commutes with $\A_R$, since $\A_L$ and $\A_R$ commute.
  In section \ref{product}, we will understand more precisely that $\A_R$, as described in eqn. (\ref{nalign}), is a factor of Type II$_\infty$.   
  
  We have expressed eqn. (\ref{nalign})  in terms of the boundary algebras.   However, duality tells us to identify these boundary algebras with bulk algebras
  $\A_r$ and $\A_\ell$:
  \be\label{duality} \A_R=\A_r,~~~\A_L=\A_\ell. \ee
  Here $\A_r$ is a bulk algebra of observables on the right side of the black hole
  that incorporates $\A_{r,0}$, the observable $\U$ that is central at $N=\infty$, and $1/N$ corrections, and 
  $\A_\ell$ is an analogous algebra on the left side of the black hole.
  
  One important fact about eqn. (\ref{nalign}) is that it implies that to all orders in $1/N$, the algebras $\A_L$ and $\A_R$ have a group of outer automorphisms
  that is not part of the exact theory.    This is simply the abelian group of translations $U\to U+c$, $c\in \R$, generated by $\Pi$.   To see that this is
  an outer automorphism of $\A_R$, observe that since the constant $c$ is anyway part of the algebra $\A_{r,0}$, it does not matter whether we adjoin
  $\U+\h h/\beta N$ or $\U+c+ \h h/\beta N$ to $\A_{r,0}$.   Similarly, since $c$ is anyway part of $e^{\i\Pi \h h/\beta N} \A_{\ell,0} e^{-\i \Pi \h h/\beta N}$,
  it does not matter if we adjoin $\U$ or $\U+c$ to that algebra.   It will turn out that there is a notion of entropy for the algebra $\A_L$ or $\A_R$ -- in contrast to
  $\A_{r,0}$ and $\A_{\ell,0}$, for example --  but this
  entropy is shifted by a constant if $\U$ is shifted.   So the outer automorphism will mean that perturbatively in $1/N$, one can only define entropy differences.

An analog of eqns. (\ref{idgo}) and (\ref{widgo}) was expressed by Jafferis, Lewkowycz, Maldacena, and Suh (JLMS) \cite{JLMS}
 in terms of the ``modular Hamiltonian'' of the boundary theory, which is defined
as minus the logarithm of the density matrix.
The density matrix of the theory on the right boundary is $\rho_R=\frac{1}{Z}e^{-\beta H_R}$, so the modular Hamiltonian is $-\log \rho_R=\beta H_R+\log Z$.   
A similar formula holds for $\rho_L$.  Here we view
$\log Z$ as a $c$-number function of $N$; in fact, $\log Z=-\beta\la H_R\ra+\la S\ra$, where $\la S\ra $ is the thermal entropy at inverse temperature $\beta$, 
also viewed as a $c$-number function of $N$.
The JLMS  formula for $\log\,\rho_R$ and $\log\,\rho_L$, which  has been important in understanding entanglement wedge reconstruction, reads 
 \begin{align}\label{oldone} \beta H_R+\log Z& = h_r +\frac{A}{4G}\cr
 \beta H_L +\log Z &= h_\ell +\frac{A}{4G}, \end{align}
 where $A$ is the area of the horizon, and a formal splitting $\h h=h_r-h_\ell$ is used.  
 Equivalently,
 \begin{align}\label{nago}\beta H_R'& = h_r +\frac{A}{4G}-\la S\ra \cr
                     \beta H_L'& =h_\ell +\frac{A}{4G}-\la S\ra .\end{align}
A comparison to eqn. (\ref{widgo}) shows that the relation is
\be\label{polin} \beta N \U+\la S\ra = h_\ell +\frac{A}{4G}.\ee

It is believed that only two linear combinations of the three operators $h_r,$ $h_\ell$, and $A/4G$, namely $h_r+A/4G$ and $h_\ell+A/4G$ (or equivalently $\h h$
and $\U$) 
 are actually well-defined quantum mechanically.    This statement is related to the observation by Susskind and Uglum \cite{SU}
 that the generalized entropy of a black hole, namely
 $A/4G+S_\out$ (where $S_\out$ is the entropy outside the horizon) is better-defined than either term is separately.
 If one believed that $h_\ell$, $h_r$, and $A/4G$ are all separately well-defined, one would  likely conclude that $h_\ell$ and $h_r$, being defined
 by integrals on the left and right of the horizon, are respectively elements
 of $\A_\ell$ and $\A_r$ and therefore of $\A_L$ and $\A_R$.   Since $H'_L$ and $H'_R$  are also contained in $\A_L$ and $\A_R$, one would
 then deduce that $A/4G$ is contained in both algebras and therefore is central in each.   This is actually true to lowest order in $G$, but as we have seen
 not beyond lowest order.

  In sections \ref{product}-\ref{entropy}, we will explain some of the mathematical theory of the algebras that  we encountered in eqn. (\ref{nalign}).
These are algebras of a special type that has been important in the mathematical theory of von Neumann algebras of Type III.    In explaining the theory,
it is convenient to assume that we are working with ordinary functions of $N$ rather than formal power series in $1/N$.  If we were working with ordinary
functions of $N$, it would not matter whether what we adjoin to $\A_{r,0}$ to build $\A_R$ is $\U+\h h/\beta N$ or $\beta N(\U+\h h/\beta N)=\beta N \U+\h h$.
So, setting 
\be\label{defx} X=\beta N\U,\ee 
 we restate eqn (\ref{nalign}) in the form
\be\label{palign} \A_R=\A_{r,0}\rtimes \A_{\h h+X}.\ee
However, for the application to gravity, all formulas of  sections \ref{product}-\ref{entropy} are  supposed to be expanded in a formal power series in $1/N$
around $N=\infty$, since that was the setting for the derivation of eqn. (\ref{nalign}).

\subsection{The Crossed Product}\label{product}

Suppose that a von Neumann algebra  $\A$ acts on a Hilbert space $\H$, and let $T$ be a self-adjoint operator on $\H$
that generates a group of automorphisms of $\A$.   The condition for this is that 
\be\label{toddo}e^{\i T s}\a e^{-\i T s}\in \A, ~~~~~\forall ~\a\in \A,~s\in \R.\ee
This gives an action of the additive group $\R$ on $\A$ by automorphisms.  (This action is not always faithful since in general some elements of $\R$ may act trivially.)
The ``crossed product'' of $\A$ by $\R$, sometimes denoted $\A\rtimes \R$, is defined as follows.   Let $L^2(\R)$ be the space of square-integrable functions of
a real variable $X$.  Then $\A\rtimes\R$ is the algebra that acts on $\H\otimes L^2(\R)$ and is obtained by adjoining $T+X$, or more precisely bounded functions
of $T+X$, to $\A$.   Thus $\A\rtimes\R$ is generated, as an algebra, by operators $\a\otimes 1,\,\a\in \A$ and $e^{\i s T}\otimes e^{\i s X}$, $s\in \R$.  Additively, $\A\rtimes\R$ is generated by operators $\a e^{\i s T}\otimes e^{\i s X}$, $\a\in \A$, $s\in\R$.   To see that operators of the form  $\a e^{\i s T}\otimes e^{\i s X}$
do form an algebra, observe that for $\a,\b\in\A$, $s,t\in\R$, we have
\be\label{weggo}(\a \,e^{\i s T}\otimes e^{\i s X} )(\b\, e^{\i t T}\otimes e^{\i t X})=\a \,e^{\i s T}\b e^{-\i s T} \,e^{\i (s+t)T} \otimes e^{\i(s+t)X}, \ee
and since eqn. (\ref{toddo}) tells us that $e^{\i s T}\b e^{-\i s T}\in\A$, the right hand side is in $\A\rtimes\R$.
One can often omit the tensor product symbol in formulas such as those of this paragraph without causing confusion.

The automorphism group that we have considered is said to be inner if $e^{\i s T}\in \A$, $s\in\R$; otherwise it is outer.  
In the case of an inner automorphism group,  adjoining $T+X$ to $\A$ is equivalent to just adjoining $X$,  since bounded functions of $T$ are already part of $\A$.
So in this case,  $\A\rtimes \R$ is just a tensor product $\A\otimes \A_X$,
where $\A_X$ is the commutative algebra of bounded functions of $X$. So  the crossed product construction applied to an inner automophism
group does not give anything essentially new, and it certainly does not give a factor, since it gives an algebra with the large center $\A_X$.
  All automorphisms of a Type I factor are inner, but that is not true for Type II or Type III.   The crossed product by $\R$ is therefore 
  potentially most interesting for Type II or Type III.

For the example that is important in the present article, let  $\Psi\in\H $ be a cyclic separating vector for $\A$, and let $\Delta=e^{-\h h}$ be the 
corresponding modular operator.   The main theorem of Tomita-Takesaki theory says (in part) that $\Delta^{-\i s}$ is an automorphism of $\A$; in other words, with
$T=\h h$,
eqn. (\ref{toddo}) is satisfied.   For an algebra of Type I, to prove this is a simple exercise with density matrices.   In general, the proof is not so simple, but
there is a relatively simple proof \cite{Longo} in the case of most importance in physics, which is
a hyperfinite algebra (an algebra that can be approximated by  finite-dimensional matrix algebras).
The automorphism group $\R_\Psi$  generated by $\h h$ is  called the modular automorphism group.  
Evidently, a restatement of eqn. (\ref{palign}) 
 is that $\A_R=\A_{r,0}\rtimes \R_{\Psi_\TFD}$.   
 
 In the mathematical theory of Type III$_1$ factors  \cite{Takesaki,Connes,ConnesBook},
 an important fact is that the crossed product of such a factor by its modular automorphism
 group $\R_\Psi$ (for any $\Psi$) is a factor of Type II$_\infty$.  This result is relevant for us, because $\A_{r,0}$ is a  Type III$_1$ factor,
 and therefore $\A_R$ is a Type II$_\infty$ factor.    In particular, $\A_R$ is a factor, and thus has trivial center, as was claimed in section \ref{beyond}.
The Type II nature of $\A_R$ will ultimately enable us to define traces, density matrices, and entropies.    A Type III$_1$ factor is actually the only type of von
Neumann algebra whose crossed product with its modular automorphism group is a factor.   So it is only because $\A_{r,0}$ is of Type III$_1$ that $\A_R$ is
a factor.

In the mathematical theory of Type III$_1$ algebras, it is essential 
 that, despite appearances, for any von Neumann algebra $\A$ with cyclic separating vector $\Psi$,  the algebra 
 $\A\rtimes \R_\Psi$ does not depend on\footnote{By contrast, if one more naively omits $X$ and simply extends the algebra
 $\A$ by adjoining bounded functions of $\h h_\Psi$ (as opposed to functions of $\h h_\Psi+X$), the resulting algebra does depend on the choice of $\Psi$.}   $\Psi$, up to a unique equivalence.   This means that when one
studies $\A\rtimes \R_\Psi$, one is learning about properties of $\A$, not about properties of the pair $\A,\Psi$.   Physically,  the fact
that $\A\rtimes\R_\Psi$  is essentially  independent of $\Psi$  is somewhat similar to a statement of background independence.\footnote{To achieve full
background independence, one would want to define the algebra in a way that is unchanged if $\Psi$ is replaced by any other state.   Here, we have the more
modest result that $\Psi$ can be replaced by any other state in the thermofield double Hilbert space at a particular temperature.   We lost the chance for full
background independence when we replaced single-trace operators $\Tr\,W$ by subtracted versions $\W=\Tr\,W-\la \Tr\,W\ra$ that have a large $N$ limit.  The subtraction depended on the temperature, and at this stage the chance for full background independence was lost.   As a step toward full background independence,
one could avoid this subtraction and consider operators that are allowed to grow as a power of $N$ for $N\to\infty$.   This appears to present other technical
difficulties.}   We will call it state-independence.
State-independence of  $\A\rtimes\R_\Psi$  is proved using the Connes cocycle \cite{Connes}, which has had a number of interesting
recent applications to quantum field theory and gravity \cite{FC,Lashkari,Bousso,BoussoTwo}. 

As a first step toward the explanation, recall that 
if $\A$ acts on $\H$ with cyclic separating vector $\Omega$, the Tomita operator $S_\Omega$ is  defined by\footnote{To be more precise, $S_\Omega$ is the closure
of the operator defined by the following condition. The same remark applies for $S_{\Psi|\Omega}$ below, and similarly later to the Tomita operators of the commutant
$\A'$.  Some basic background on the Tomita operators and related matters can be found, for example, in \cite{OlderLectures}.   In the present article,
we use the most common convention for the relative Tomita and modular operators.   In \cite{OlderLectures}, a different choice was made for a reason explained in footnote
16 of that article.}  
$S_\Omega\a\Omega=\a^\dagger\Omega,~~\a\in\A.$
   One also defines the modular operator $\Delta_\Omega=S_\Omega{}^\dagger S_\Omega$.    More generally, if $\Psi\in\H$ is another vector (which we will usually
   assume to be also cyclic separating),
one defines the relative Tomita operator $S_{\Psi|\Omega}$ via the condition 
$S_{\Psi|\Omega}\a\Omega=\a^\dagger\Psi$.   The relative modular operator is defined to be
$\Delta_{\Psi|\Omega}=S_{\Psi|\Omega}^\dagger S_{\Psi|\Omega}$.   Thus in particular 
\be\label{zombo} S_{\Psi|\Psi}=S_\Psi,~~~ \Delta_{\Psi|\Psi}=\Delta_\Psi.\ee

  The Connes cocycle is defined as
\be\label{boxo} u_{\Psi|\Omega}(t) =\Delta_{\Psi|\Omega}^{\i t}  \Delta_{\Omega} ^{-\i t}  =\Delta_\Psi^{\i t}\Delta_{\Omega|\Psi}^{-\i t}  \ee
Clearly $u_{\Psi|\Omega}(t)$ is unitary for real $t$, and $u_{\Psi|\Psi}(t)=1$.
Some important  additional properties of $u_{\Psi|\Omega}(t)$ are:
\begin{itemize}
\item{$u_{\Psi|\Omega}(t)\in \A.$}
\item{The two formulas for $u_{\Psi|\Omega}(t)$ given in eqn. (\ref{boxo}) are in fact equal.}
\item{If $\Phi$, $\Omega$ and $\Psi$ are three cyclic separating states, then $u$ satisfies a chain rule 
\be\label{chain}u_{\Psi|\Omega}(t)u_{\Omega|\Phi}(t) =u_{\Psi|\Phi}(t) \ee 
and in particular
\be\label{lotto}u_{\Psi|\Omega}(t) u_{\Omega|\Psi}(t)=1. \ee 
The equivalence of the two formulas in eqn. (\ref{boxo})  actually amounts  to this special case of the chain rule.
} \end{itemize}
The proofs of these statements (due originally to Connes \cite{Connes}) are explained in section 6 of \cite{Lashkari}.

Let $\A'$ be the commutant of $\A$  (the algebra of bounded operators on $\H$ that commute with $\A$).   If a vector is cyclic and separating for $\A$, then  it is also cyclic
and separating for $\A'$, so one can define operators $S'_\Psi$, $S'_{\Psi|\Omega}$
 characterized by $S'_\Psi \a' \Omega=\a'{}^\dagger\Omega$, $S'_{\Psi|\Omega}\a'\Omega=\a'{}^\dagger
\Psi$.   Likewise, one defines $\Delta'_\Omega=S_\Omega'{}^\dagger S'_\Omega$, $\Delta'_{\Psi|\Omega}=S'_{\Psi|\Omega}{}^\dagger S'_{\Psi|\Omega}$.
The Connes cocycle for $\A'$ is defined by the obvious analog of eqn. (\ref{boxo}).    However, one has $\Delta'_\Omega=\Delta_\Omega^{-1}$,
$\Delta'_{\Psi|\Omega}=\Delta_{\Omega|\Psi}^{-1}$.   So one can write the cocycle for $\A'$ in terms of the modular operators for $\A$:
\be\label{noxo} u'_{\Psi|\Omega}(t) =\Delta_{\Omega|\Psi}^{-\i t}  \Delta_{\Omega} ^{\i t}  =\Delta_\Psi^{-\i t}\Delta_{\Psi|\Omega}^{\i t}  \ee
One has the same three properties as before: $u'_{\Psi|\Omega}\in\A'$; the two expressions for $u'_{\Psi|\Omega}$ are equal; given three states,
$u'$ satisfies a chain rule $u'_{\Psi|\Omega}(t)u'_{\Omega|\Phi}(t) =u'_{\Psi|\Phi}(t).$

We now want to show that, given any two cyclic separating vectors $\Omega,\Psi$, the two algebras $\A\rtimes\R_\Omega$ and $\A\rtimes\R_\Psi$
are conjugate in a natural fashion.  To be more precise, setting $P=-\i \d/\d X$, the algebras are conjugate via the operator $u'_{\Psi|\Omega}(P)$ that we get by substituting $t$ with $P$ in the definition of
$u'_{\Psi|\Omega}(t)$.   (This substitution makes sense because $P$ commutes with $u'_{\Psi|\Omega}(t)$ for all $t$, so we can define an operator 
$u'_{\Psi|\Omega}(P)$ that acts on a state with $P=t$ as $u'_{\Psi|\Omega}(t)$.)   Thus, the claim is that
\be\label{thec} \A\rtimes\R_\Psi =u'_{\Psi|\Omega}(P) \bigl(\A\rtimes \R_\Omega\bigr) u'_{\Psi|\Omega}(P)^\dagger.\ee
The chain rule implies that this construction leads to a unique conjugacy between the crossed product algebras for any two states; in other words,
given three states $\Phi,\Omega,\Psi$, it does not matter if we conjugate directly from  $\A\rtimes\R_\Phi$  to
$ \A\rtimes\R_\Psi$ according to eqn. (\ref{thec}), or if we conjugate first from $\A\rtimes\R_\Phi$ to  $\A\rtimes\R_\Omega$ and then to $ \A\rtimes\R_\Psi$.  This ability  to select a {\it unique}
conjugacy between the two algebras means that they are really physically the same rather than just being abstractly isomorphic.\footnote{Abstract isomorphism
would have very little import.  For example,  in any quantum field theory, the algebra of observables 
in  any topologically simple local region in any spacetime is believed to be a factor of Type III$_1$,
so all these are isomorphic; but there is no useful sense in which they are ``the same,'' since there is no distinguished isomorphism between them.}

Since $u'_{\Psi|\Omega}$ is invertible, with inverse $u'_{\Omega|\Psi}$, to establish the conjugacy claimed in eqn. (\ref{thec}), one just needs
to show that $u'_{\Psi|\Omega}$ conjugates any of the generators of $\A\rtimes \R_\Omega$ to an element of $\A\rtimes\R_\Psi$.
We will use the fact that $\A\rtimes\R_\Omega$ is generated by operators $\a\otimes 1$, $\a\in \A$, along with $e^{\i \h h_\Omega s}\otimes e^{\i X s}=\Delta_\Psi^{-\i s}
\otimes e^{\i  Xs }$, $s\in \R$.
More informally, the generators are $\a$ and $\Delta_\Omega^{-\i s}e^{\i Xs}$.     Similarly,  $\A\rtimes \R_\Psi$ is generated by operators 
$\a$ and $\Delta_\Psi^{-\i s}e^{\i X s}$.    The conjugacy we want is trivial for the generators $\a\in \A$: since $u'_{\Psi|\Omega}(s)\in \A'$ for all $s$, and $\A$
commutes with both $\A'$ and $P$, $u'_{\Psi|\Omega}(P)$ commutes with $\A$.
   So we just need to show the desired conjugacy for the other generators $\Delta_\Omega^{-\i s}e^{\i sX}$:
\be\label{doffo} u'_{\Psi|\Omega}(P) \Delta_\Omega^{-\i s}e^{\i s X}u'_{\Psi|\Omega}(P)^\dagger\in \A\rtimes \R_\Psi,~~~\forall s\in\R.\ee
Since $[X,P]=\i$, we have $e^{\i s X}u'_{\Psi|\Omega}(P)^\dagger= u'_{\Psi|\Omega}(P-s)^\dagger e^{\i s X}$.
So
\begin{align}\label{finally} u'_{\Psi|\Omega}(P) \Delta_\Omega^{-\i s}e^{\i s X}u'_{\Psi|\Omega}(P)^\dagger &=u'_{\Psi|\Omega}(P) \Delta_\Omega^{-\i s}u'_{\Psi|\Omega}(P-s)^\dagger e^{\i s X} \cr & = \Delta_{\Omega|\Psi}^{-\i P} \Delta_\Omega^{\i P} \Delta_\Omega^{-\i s} \Delta_\Omega^{-\i P+\i s}\Delta_{\Omega|\Psi}
^{\i P-\i s} e^{\i s X} \cr &=\Delta_{\Omega|\Psi}^{-\i s}e^{\i s X} = \Delta_\Psi^{-\i s} e^{\i s X} u_{\Psi|\Omega}(s)  .  \end{align}
Since  $ u_{\Psi|\Omega}(s)\in \A\subset \A\rtimes \R_\Psi$ and $\Delta_\Psi^{-\i s} e^{\i s X}$ is also a generator of $\A\rtimes \R_\Psi$,
the final expression in eqn. (\ref{finally}) is the product of two elements of $\A\rtimes \R_\Psi$, and therefore is an element of this algebra.   
This completes the proof that $u'_{\Psi|\Omega}(P)$ conjugates $\A\rtimes \R_\Omega$ to $\A\rtimes \R_\Psi$.

\subsection{Classical-Quantum States And The Modular Operator}\label{cqm}

In section \ref{review}, we encountered an algebra $\A_{r,0}\otimes \A_\U$ with a large center. Because of this center, every state for
this algebra is classical-quantum in a natural sense.  For  the crossed product algebra 
$\A_R=\A_{r,0}\rtimes \R_\Psi$ acting on $\h\H=\H\otimes L^2(\R)$, that is no longer true, but a reasonable class of classical-quantum states
are states of the form $\chi\otimes g^{1/2}(X)$, $\chi\in\H$, $g^{1/2}(X)\in L^2(\R)$.  
It follows from the analysis of state independence in section \ref{product} that even though the algebra $\A_{r,0}\rtimes \R_\Psi$ is state-independent,
this  notion of a classical-quantum state is state-dependent.   When we identify $\A_{r,0}\rtimes \R_\Omega$ with $\A_{r,0}\rtimes \R_\Psi$
by conjugating the algebra of observables by $u'_{\Psi|\Omega}(P)$, we must also act on the states with the same operator $u'_{\Psi|\Omega}(P)$.
But $u'_{\Psi|\Omega}(P)$ does not map a classical-quantum state of the class just described to a state of the same class.

It  turns out to be useful to understand modular theory for classical-quantum states of the special form $\h\Psi=\Psi\otimes g^{1/2}(X)$, where the quantum
part of $\h\Psi$ is the same state $\Psi$ that was used to define the crossed-product.   Thus, given the modular operator $\Delta_\Psi$
for a state $\Psi$ of an algebra $\A$, there is a simple and informative formula for the modular operator $\h \Delta_{\h \Psi}$ of 
$\h\Psi$.   We assume here that $g^{1/2}(X)$ is everywhere positive,
which ensures that $\h\Psi$ is  cyclic separating.   

We will use the 
following   characterization of the modular operator $\Delta_\Psi$ of an algebra $\A$ with cyclic separating state $\Psi$. 
$\Delta_\Psi$ is characterized by the fact that, for all $\a,\b\in \A$,
\be\label{zolbov} \la\Psi|\a\b|\Psi\ra =\la\Psi|\b \Delta_\Psi \a|\Psi\ra. \ee      To verify eqn. (\ref{zolbov}), one computes \begin{align}\label{patience} \hspace{-.6cm}
\la\Psi|\b \Delta_\Psi\a|\Psi\ra&=\la\Psi|\b S_\Psi{}^\dagger S_\Psi\a|\Psi\ra=\la\b^\dagger\Psi| S_\Psi{}^\dagger |S_\Psi\a\Psi\ra\cr& =
\la  S_\Psi\a\Psi|S_\Psi| \b^\dagger\Psi\ra=\la\a^\dagger\Psi|\b\Psi\ra=\la\Psi|\a\b|\Psi\ra.\end{align}   We used the definition of the adjoint of
an antilinear operator, $\la\Psi|S_\Psi^\dagger|\Psi'\ra=\la\Psi'|S_\Psi|\Psi\ra$.  
We will also need the KMS condition.  Set $\h h_\Psi=-\log \Delta_\Psi$, and for $\a\in \A$, let $\a_u=e^{\i\h h_\Psi u} \a e^{-\i \h h_\Psi u}$.
The KMS condition is a property of functions such as $\la \a_u \b\ra=\la\Psi|\a_u \b |\Psi\ra$; if $\Psi$ is a thermofield double
state, then $\a_u$ is a time translate of $\a$ and these are real time thermal correlation functions.
Replacing $\a$ with $\a_u$ in eqn. (\ref{zolbov})  we get the KMS condition
\be\label{kms}\la\Psi|\a_u\b|\Psi\ra=\la\Psi|\b \a_{u+\i} |\Psi\ra. \ee
We used  $\Delta_\Psi\Psi=\Psi$ to get $\Delta_\Psi \a_u|\Psi\ra=\Delta_\Psi^{\i s} \a_u \Delta_\Psi^{-\i s}| \Psi\ra$ with $s=-\i$, and we interpreted this as $\a_{u+\i}\Psi$.
This explanation has been rather cavalier; see for example section 4.2 of \cite{OlderLectures} for more detail.  A precise statement of the KMS condition  is that the function $\la\Psi|\a_u \b|\Psi\ra$,
originally defined for real $u$, analytically continues to a function holomorphic in the strip $-1\leq \Im\,u\leq 0$, and the function $\la\Psi|\b\a_u|\Psi\ra$,
likewise initially defined for real $u$,  analytically continues to a function holomorphic in the strip $0\leq \Im\,u\leq 1$, in such a way that eqn. (\ref{kms}) holds. 
Because $\h h_\Psi\Psi=0$, these functions  satisfy
\be\label{wms}\la\Psi|\a_u\b_v|\Psi\ra=\la\Psi|\a_{u+s}\b_{v+s}|\Psi\ra.\ee
In the context of the thermofield double, this is time translation symmetry.

We want to find the operator $\h\Delta_{\h\Psi}$ that obeys
\be\label{toxo} \la \h\Psi|\h\a \h \b |\h\Psi\ra=\la\h\Psi|\h \b \h\Delta_{\h\Psi}\h a|\h\Psi\ra,~~ \forall~ \h a,\h b\in \A\rtimes\R_\Psi.\ee
It turns out that \be\label{turnco} \h\Delta_{\h\Psi}=\Delta_\Psi\cdot  g(\h h_\Psi+X)(g(X))^{-1}.\ee   In proving this, it is convenient to introduce the Fourier transform of $g(X)$:
\begin{align}\t g(w)& =\int_{-\infty}^\infty \d X \,e^{\i w X} g(X) \cr
                        g(X)&=\frac{1}{2\pi}\int_{-\infty}^\infty \d w\,e^{-\i wX}\t g(w). \end{align}
So another way to write the formula  for $\h\Delta_{\h\Psi} $ is
\be\label{nother} \h\Delta_{\h\Psi}=  \frac{\Delta_\Psi}{2\pi g(X)}\int_{-\infty}^\infty \d w \,e^{-\i w (\h h_\Psi+X)} \t g(w)        .\ee      
The operators $\Delta_\Psi$, $\h h_\Psi$, and $X$ all commute, so there are no subtleties of operator ordering.                 

It is enough to verify (\ref{toxo})  for operators that form an additive basis of $\A\rtimes \R_\Psi$, so we can take $\h\a=\a e^{\i s(\h h_\Psi+X)}$,
$\h \b =\b e^{\i t(\h h_\Psi+X)}$, with $\a,\b\in\A$ and $s,t\in\R$.  
We have
\begin{align} \label{zzz}\la\h\Psi|\h \a\h \b|\h\Psi\ra&=\int_{-\infty}^\infty \d X g(X) \la\Psi|\a e^{\i s(\h h_\Psi+X)} \b e^{\i t(\h h_\Psi+X)}|\Psi\ra
\cr &=\int_{-\infty}^\infty \d X g(X) e^{\i(s+t)X} \la\Psi|\a \b_s|\Psi\ra =\t g(s+t)  \la\Psi|\a \b_s|\Psi\ra.\end{align}
On the other hand, eqn. (\ref{nother}) leads to 
\begin{align}\label{ohmy}\la\h\Psi|\h\b \h\Delta_{\h\Psi}\h \a|\h\Psi\ra&=\int_{-\infty}^\infty \d X \frac{1}{2\pi}\int_{-\infty}^\infty \d w \,\t g(w)\la\Psi |
\b e^{\i t(\h h_\Psi+X) }\Delta_\Psi e^{-\i w(\h h_\Psi+X)}\a e^{\i s(\h h_\Psi+X)}|\Psi\ra\cr&
=\frac{1}{2\pi}\int_{-\infty}^\infty \d w\,\t g(w) \int_{-\infty}^\infty \d X \,e^{\i(s+t-w)X}\la\Psi|\b \a_{t-w+\i} |\Psi\ra\cr
&=\t g(s+t)\la\Psi|\b\a_{\i-s}|\Psi\ra=\t g(s+t)\la\Psi|\a_{-s}\b|\Psi\ra\cr&=\t g(s+t)\la\Psi|\a\b_s|\Psi\ra. \end{align}
We integrated over $X$ and $w$ and used the KMS condition and time translation symmetry.  This completes the proof of the formula for
$\h\Delta_{\h\Psi}$.

Eqn. (\ref{turnco}) implies  that we can factor $\h\Delta_{\h\Psi}$ as
\be\label{arno}\h \Delta_{\h\Psi}=\t K K, \ee with
\begin{align}\label{narno}  K&= \Delta e^{-X} g(\h h_\Psi+X)=e^{-(\h h_\Psi+X)} g(\h h_\Psi+X) \cr
     \t K& =\frac{e^{X}}{ g(X)}. \end{align}
     The point of this factorization is as follows.
  Since it is only a function of $\h h_\Psi+X$, $K$ is contained in the crossed product algebra $\A\rtimes \R_\Psi$.
  (This statement assumes that the function $g(X)$ has the property that $g(X) e^{-X}$ is bounded; otherwise we should say
  that $K$ is affiliated to $\A\rtimes \R_\Psi$, meaning that bounded functions of $K$ are in $\A\rtimes\R_\Psi$.)   Likewise
  $\t K$ is a function only of $X$, so it commutes with $\A\rtimes\R_\Psi$ and belongs to the commutant $(\A\rtimes \R_\Psi)'$.   
  
  Existence of such a  factorization means that the modular automorphism group of $\h\Psi$ is inner.  Indeed for $\a\in\A\rtimes \R_\Psi$,  $\h\Delta_{\h\Psi}^{\i s}\a
  \h\Delta_{\h\Psi}^{-\i s}=K^{\i s}\a K^{-\i s}$, because $\a$ commutes with $\tilde K$. Since $K\in \A\rtimes \R_\Psi$, the automorphism $\a\to K^{\i s}\a K^{-\i s}$ is inner.
For an algebra of Type III, the modular automorphism group is never inner.   So for any $\A$, the crossed product algebra $\A\rtimes \R_\Psi$ is always
of Type I or Type II.    

A noteworthy fact is that the factorization in eqn. (\ref{arno}) is not quite unique.   If we shift $X\to X+c$, we get
\be\label{bildo} K\to e^{-c }K,~~~~~\t K\to e^c\t K. \ee         The group of constant shifts of $X$, $X\to X+c$, is an outer automorphism
group of the crossed product algebra; it was already introduced in 
 section \ref{beyond}.     
  There is no further indeterminacy in the factorization of $\h\Delta$ 
provided the crossed product algebra is a factor.       As already remarked, this happens precisely in the case of interest to us that $\A$ is of Type III$_1$. 

\subsection{Traces} \label{traces}

A trace on an algebra  is a linear function on the algebra such that $\Tr\,\a\b=\Tr\,\b\a$ for any two elements $\a,\b$ in the algebra. 
For example, for any positive integer  $N$, the algebra of $N\times N$ matrices  has such a trace.
A more subtle case is the Type I$_\infty$  von Neumann algebra  of all bounded operators on a separable (infinite-dimensional)
Hilbert space $\H$.  This algebra has a trace, but it is not
defined for all elements of the algebra, only for a subalgebra consisting of operators
 that are ``trace class.'' For example, the trace of the identity element in an algebra of Type I$_\infty$ is not defined; it would be $+\infty$.

The factorization  of the modular operator for a classical-quantum state $\h\Psi$ 
leads immediately to the existence of a trace, albeit in general one that is defined only for a subalgebra.
For $\h\a\in \A\rtimes \R_\psi$, one defines 
\be\label{donkeyd}\Tr\,\h\a=\la\h\Psi|\h\a K^{-1}|\h\Psi\ra. \ee
Making use of eqn. (\ref{zolbov}), we get
\be\label{onkey} \Tr\,\h\a\h\b =\la\h\Psi|\h\a\h\b K^{-1}|\h\Psi\ra=\la \h\Psi|\h\b K^{-1} \h\Delta_{\h\Psi} \h\a|\h\Psi\ra.\ee
Since $\h\Psi=\h\Delta_{\h\Psi}^{-1}\h\Psi$, the right hand side of eqn. (\ref{onkey}) is $\la\h\Psi|\h\b K^{-1} \h\Delta_{\h\Psi} \h\a\h\Delta_{\h\Psi}^{-1}|\h\Psi\ra$.
Writing $\h\Delta_{ \h \Psi}=K\t K$, where $\t K$ commutes with $K$ and with $\h\a$, we get
\be\label{nkey}\Tr\,\h\a\h\b=\la\h\Psi|\h\b K^{-1}K \h\a K^{-1}|\h\Psi\ra=\la\h\Psi|\h\b\h\a K^{-1}|\h\Psi\ra = \Tr\,\h\b\h\a, \ee
completing the proof that this function is a trace.  

Using eqn. (\ref{narno}) for $K$, along with $\Delta_\Psi\Psi=\Psi$, $\h h_\Psi|\Psi\ra=0$, the definition of the trace can be simplified to
\be\label{hockey} \Tr\,\h\a =\left\la\h\Psi\left|\h\a \frac{e^X}{g(X)}\right|\h\Psi\right\ra=\int_{-\infty}^\infty\d X \,e^X \la\Psi|\h\a |\Psi\ra. \ee
In writing the last formula, we view $\h \a$ as a function of $X$ with values in operators on $\H$, so that the matrix element $ \la\Psi|\h\a |\Psi\ra$
is a function of $X$, which goes into the integral over $X$.  This formula shows
that the trace on $\A\rtimes \R_\Psi$ is not defined for all elements of this algebra, since the integral over $X$ may not converge.   For example,
it does not converge if $\h\a$ is an element of the original algebra $\A$.  

 A noteworthy fact is that in the definition of the trace, the dependence on the function $g(X)$ has disappeared.
In other words, we used the function $g(X)$ to define a classical-quantum state that in turn motivated the formula for the trace, but the trace
that we have defined does not depend on the choice of $g(X)$.  That does not mean that the definition of the trace is completely canonical.
The outer automorphism of the crossed product that shifts $X$ to $X+c$
has a simple action on the trace:
\be\label{hugge}\Tr\,\h \a\to e^c\Tr\,\h \a.\ee
It is no accident, of course, that the only indeterminacy of the trace is
 a simple rescaling.  When a von Neumann algebra is a factor (its center consists only of complex scalars),
a trace on this algebra, if it exists, is unique up to a scalar multiple.   

Among those von Neumann algebras that are factors, Type III algebras do not admit any trace, and Type I and Type II$_1$ algebras admit a trace that  is not rescaled by any outer automorphism.   The only case of a factor that admits a trace that is rescaled by an outer
automorphism is a factor of\footnote{See \cite{Lecture}, section 3.6, for another explanation of why a Type II$_\infty$ factor has a trace whose
normalization cannot be canonically determined.}  Type II$_\infty$.   This is the case of primary interest in the present article.

We can use eqn. (\ref{hockey}) to get a useful formula for the trace for elements of a certain subalgebra of $\A\rtimes \R_\Psi$.   We recall that $\A\rtimes \R_\Psi$
contains elements $\a e^{\i\u(\h h_\Psi+X)}$ for any $u\in\R$, and likewise contains linear combinations $\sum_i \a_i e^{\i\u_i(\h h_\Psi+X)}$.
Taking limits, $\A\rtimes \R_\Psi$ contains operators of the form  $\h\a=\int_{-\infty}^\infty \d u \,\a(u)e^{\i u(\h h_\Psi+X)}$ with a smooth function $\a(u)$ that vanishes
for large $u$.    Let us impose a further condition that
the function $\a(u)$ is holomorphic for $0\leq {\mathrm{Im}}\,u\leq 1$.    In this case, we can get a convenient formula for $\Tr\,\h\a$:
\be\label{zilbo}\Tr\,\h\a=\int_{-\infty}^\infty\d X \,e^X\la\h\Psi|\h\a |\h\Psi\ra=\int_{-\infty}^\infty \d X\d u\, e^{X+\i u X}\la\Psi|\a(u)|\Psi\ra. \ee
After shifting the contour in the $u$ integral from $\R$ to $\i+\R$ and integrating first over $X$ to get a delta function in $u$, we find
\be\label{nilbo}\Tr\,\h\a =2\pi  \la\Psi|\a(\i)|\Psi\ra. \ee

The condition for $\a(u)$ to be holomorphic in the strip is rather special, but elements of $\A\rtimes \R_\Psi$ associated with functions that have that
property form a subalgebra and it is interesting to check the cyclic property of the trace for elements of this subalgebra.   Suppose that
 $\h\a=\int_{-\infty}^\infty \d u \a(u)e^{\i u(\h h_\Psi+X)},$  $\h\b=\int_{-\infty}^\infty \d u \b(u)e^{\i u(\h h_\Psi+X)},$ where $\a(u)$ and $\b(u)$ are both
 holomorphic in the strip.  Then  defining $\b_u(v)=e^{\i \h h_\Psi u} \b(v)e^{-\i \h h_\Psi u}$, we get 
 \be\label{zimbo} \h \a\h \b=\int_{-\infty}^\infty \d u\,\d v \,\a(u) \b_u(v) e^{\i (u+v)(\h h_\Psi+X)}
 =\int_{-\infty}^\infty \d w\, \d u\, \a(u) \b_u(w-u) e^{\i w(\h h_\Psi+X)}. \ee
 So eqn. (\ref{nilbo}) gives in this case
 \be\label{ilbo} \Tr\,\h \a\h\b =2\pi \int_{-\infty}^\infty \d u\la\Psi| \a(u) \b_u(\i - u)|\Psi\ra.\ee 
 On the other hand,
 \begin{align}\label{pilbo} \Tr\,\h\b\h\a &=2\pi\int_{-\infty}^\infty \d u \la\Psi|\b(u) \a_u(\i-u)|\Psi\ra \cr
     & = 2\pi \int_{-\infty}^\infty \d u\,\la\Psi|\a_{u-\i}(\i-u) \b(u)|\Psi\ra \cr
       & = 2\pi \int_{-\infty}^\infty \d u \,\la\Psi |\a(\i-u)\b_{\i-u}(u)|\Psi\ra \cr
       & =2\pi \int_{-\infty}^\infty\d u\,\la\Psi|\a(u)\b_u(\i-u)|\Psi\ra. \end{align}
       We successively used the KMS condition (\ref{kms}), translation invariance (\ref{wms}), and a change of variables $u\to \i-u$.

\subsection{Density Matrices and Entropies}\label{entropy}

Let $\A$ be an  algebra that acts on a Hilbert space $\H$.  Suppose also that
$\A$ is equipped with a nondegenerate  trace $\Tr$ that is positive in the sense that $\Tr\,\a^\dagger\a>0$ for all nonzero $\a$.
Under these circumstances, one can define a density matrix.   The density matrix $\rho$ of a state $\Psi\in\H$ is an element $\rho\in \A$ 
such that for $\a\in\A$, $\la\Psi|\a|\Psi\ra=\Tr\,\rho\,\a$.   Such a $\rho$ will exist because of nondegeneracy of the trace.   $\rho$ will be positive
in the sense that $\Tr\,\rho\,\a^\dagger \a\geq 0$ for all $\a\in \A$.  It will satisfy $\Tr\,\rho=1$.

Such a $\rho$ is quite analogous to a density matrix in ordinary quantum mechanics.   It describes the outcomes when an arbitrary $\a\in\A$
is measured  in the state $\Psi$.

Once one has a trace and a notion of a density matrix, one can define a version of the von Neumann entropy:
\be\label{entdef}S(\rho)=-\Tr\,\rho\log\rho. \ee

In the case of the crossed product algebra $\A\rtimes \R_\Psi$, the definition of the trace is not completely canonical, since $\A\rtimes \R_\Psi$
has a group of outer automorphisms that conjugates $X$ to $X+c$, $c\in \R$.   As observed in section \ref{traces}, this rescales the trace by
$\Tr\to e^c\Tr$.   To preserve the condition $\Tr\,\rho=1$, we have to compensate by rescaling the density matrix, $\rho\to e^{-c}\rho$,
and this has the effect of shifting the entropy by the same constant $c$:
\be\label{shiftentropy} S(\rho)\to S(\rho)+c. \ee
Thus the indeterminacy in the entropy is an overall additive constant, the same for all states.  This is similar to the situation in classical physics, 
where entropy differences are defined
by relations such as $\d E = T\d S$, but there is no natural way to fix an additive constant in the entropy.

One can eliminate the arbitrary additive constant by taking the differential of eqn. (\ref{entdef}).  This gives
\be\label{nudef}\d S=-\Tr\,\d\rho \log\rho,\ee
by analogy with the classical $\d E=T\d S$ and also by analogy with what is sometimes called the first law of entanglement entropy \cite{BCHM}.
Entropy differences defined this way should be physically sensible.
 Another analogy is with the many problems, such as those studied in \cite{CH,Casini},
  in which entanglement entropy differences in quantum field theory are better defined
 than entanglement entropies.

Part of the analogy with classical physics is actually that the entropy defined in eqn. (\ref{entdef}) is not positive-definite, and in fact it is not bounded below.\footnote{This
statement involves taking literally that we are interested in the entropy of a state of a Type II$_\infty$ algebra, and not thinking in terms of an asymptotic expansion in
$1/N$.  From the standpoint of the $1/N$ expansion, one would call the entropy positive if the leading term for large $N$ is positive, regardless of the
nature of the lower order terms.   This is likely to be true in physically sensible situations.   The remarks in the text are applicable if we think of $N$ as a fixed number at which we have
a Type II$_\infty$ algebra; then, the entropy is unbounded below.  However, the analysis in this article leading to a Type II$_\infty$ algebra is really only
valid in an asymptotic expansion near $N=\infty$.}  
  The entropy of a classical harmonic oscillator
goes to $-\infty$ as the temperature goes to 0, because the system gets compressed in a smaller and smaller phase space volume.  As explained in section 3.6
of \cite{Lecture}, entropy in a Type II algebra, such as the crossed product algebra that we have encountered in black hole physics,
 can be made arbitrarily negative by disentangling qubits.

Let us compute the entropy of the classical-quantum state 
$\h\Psi =\Psi\otimes g(X)^{1/2}$ that we used to motivate the definition of the trace.   In this case, upon replacing $\h\a$ by $\h\a K$ in the definition 
$\Tr\,\h\a=\la\h\Psi|\h\a K^{-1}|\h\Psi\ra$, we get $\Tr\,\h\a K=\la\h\Psi|\h\a|\h\Psi\ra$.   So the density matrix of $\h\Psi$ is just $K$. 
From eqn. (\ref{narno}), we find  $K\log K=e^{-(\h h_\Psi+X)}g(\h h_\Psi+X)(-(\h h_\Psi+X)+\log g(\h h_\Psi +X)).$  Using eqn. (\ref{hockey}) and recalling that $\h h_\Psi|\h\Psi\ra=0$, we find
the entropy of the density matrix $K$ to be
\be\label{tungolo} S(K)=\int_{-\infty}^\infty \d X \,\left(X g(X)-g(X) \log g(X)\right). \ee
The normalization condition for the state $\h\Psi$  is
\be\label{bongo}\int_{-\infty}^\infty\d X\, g(X)=1. \ee
Clearly, subject to this normalization condition, $S(K)$ is  unbounded above and below.   If $g(X)$ is strongly peaked around a classical value $X_0$,
then $S(K)$ will be very close to $X_0$ (assuming $|X_0|\gg |\log g(X_0)|$).

In the large $N$ limit of the thermofield double state, $g(X)$ is a Gaussian, presented in eqn. (\ref{gaussf}) (recall $X=\beta N\U$).  This Gaussian
is peaked near $X=0$, not near $X=A/4G$, as we would need to get the Bekenstein-Hawking entropy.   One must recall that entropy for a Type II$_\infty$
factor involves an arbitrary additive constant, so only entropy differences can really be defined.   In effect, with the normalizations we have used, entropy is
measured relative to the classical entropy $A/4G$ of a black hole at inverse temperature $\beta$.   (Somewhat similarly, entropy in a Type II$_1$
algebra is defined relative to the entropy of a maximally entangled state.   See for example section 3.6 of \cite{Lecture}.)   If one evaluates $S(K)$ using
eqn. (\ref{gaussf}) for $g$, one finds that the $\log g$ term in eqn. (\ref{tungolo}) leads, apart from a constant,
 to a contribution $\frac{1}{2}\partial_\beta \log |\la H_R\ra|$ to $S(K)$.   This term is actually a universal logarithmic correction to the Bekenstein-Hawking
 entropy that is associated to energy fluctuations in the canonical ensemble \cite{DMB}.

We have obtained in eqn. (\ref{tungolo}) what looks like a classical formula for the entropy because of considering a special sort of classical-quantum state.
For a more generic state one would certainly not get a classical formula for the entropy.

\section{Other Conserved Charges}\label{symmetries}

In addition to time translation symmetry, the two-sided eternal black hole solution of $\N=4$ super Yang-Mills theory has additional rotational and gauge symmetries.
The symmetry group is maximized for the case that the angular momentum and conserved charges of the black hole vanish.    In this case, the rotational
symmetry group is $\Spin(4)$ and the gauge group is $\SU(4)_\Rr$, which from the boundary point of view is a group of $\Rr$-symmetries.  The global form
of the symmetry group is $G=(\Spin(4)\times \SU(4)_\Rr)/\Z_2$.    Observers on the left and right side boundaries of the black hole spacetime see separate
symmetry groups $G_L$ and $G_R$.   In the boundary theory, these symmetry groups are generated by left and right conserved charges $Q^a_L$ and $Q^a_R$,
where $a$ runs over a basis of the Lie algebra $\g$ of $G$.   

We can proceed precisely as in the discussion of time translations in section \ref{review}.  The operators $Q^a_L$ and $Q^a_R$ do not separately have
large $N$ limits in the thermofield double state $\Psi_\TFD$.   The difference\footnote{The left and right systems in the thermofield double
are CPT conjugates.  The $*$ in $Q_L^{a*}$ is CPT, which acts by complex conjugation if the group generators are defined to be hermitian
operators (or as minus complex conjugation if they are defined to be antihermitian).}    $\h Q^a=Q_R^a-Q_L^{a*}$ annihilates $\Psi_\TFD$ and does have a large $N$
limit.   The $\h Q^a$  act on the Hilbert space $\H_\TFD$ obtained by quantizing the generalized free field theory of nonzero modes that emerges in the large $N$ limit
\cite{LL,LL2} and  generate an action of the group $G$.  Let us denote the representation matrices as $W(g)$, $g\in G$.

The operators $\U^a=Q^a_R/N$
also have a large $N$ limit.   They are central in the large $N$ limit and their correlation functions are Gaussian.  To get a Hilbert space on which the $\U^a$
can act in addition to the operators of the generalized free field theory, we can extend $\H_\TFD$ to $\h\H_\TFD=\H_\TFD\otimes L^2(\g)$.  Here $L^2(\g)$ is the space
of square-integrable functions of the $\U^a$, with the $\U^a$ acting by multiplication.  (We also need to extend $\H_\TFD$ to accommodate time translations,
as described in section \ref{review}, but we omit this here.)     

This gives an adequate framework to describe the large $N$ limit, but to go to higher orders in $1/N$ in the case of a nonabelian symmetry group,
a slightly different point of view is preferable.  The global symmetry group of the boundary theory is the product $G_L\times G_R$, but the eternal black hole
solution is invariant only under a diagonal subgroup $G_D$ of $G_L\times G_R$.   This means that actually there is a family of classical solutions,
parametrized by the quotient $(G_L\times G_R)/G_D$.   This quotient is a copy of $G$; we will call that copy $G_\M$ (where $\M$ is for ``moduli space'').
Concretely, $G_\M$ parametrizes the choice of ``Wilson line'' between the left and right boundary; thus, starting with any one given solution, representing
a point that we identify with $1\in G_\M$, we make a solution corresponding to an arbitrary $g\in G_\M$ by making a gauge transformation by an element
of the gauge group (here the group of rotations and $\Rr$-symmetries) that equals 1 on the left boundary and equals $g$ on the right boundary.
The group $G_L\times G_R$ acts on $G_\M$ by $g\to g_L g g_R^{-1}$, $g_L\in G_L$, $g_R\in G_R$.   A given point $g\in G_\M$ is invariant under a
diagonal subgroup of $G_L\times G_R$ defined by $g_L g g_R^{-1}=g$ or $g_L=g g_R g^{-1}$.   We will call this group $G_g$.

For any choice of $g\in G_\M$, quantization of small fluctuations in the background of the eternal black hole gives a thermofield double Hilbert space that we will
now call $\H_{\TFD,g}$, since it depends on $g$.  Each $\H_{\TFD,g}$ is a representation of the subgroup $G_g$ of $G_L\times G_R$ that is unbroken at
the point $g\in G_\M$.
 As $g$ varies, $\H_{\TFD,g}$ varies as the fiber of a bundle $\V$ of Hilbert spaces over $G_\M$:
\be\label{zolbo}\begin{matrix} \H_{\TFD,g} & \longrightarrow & \V\cr && \downarrow \cr && G_\M.\end{matrix}  \ee
An ``improved'' thermofield double Hilbert space $\h\H_\TFD$  that incorporates the moduli in $G_\M$ can be defined as the space of $L^2$ sections of the Hilbert space bundle $\V$.    

However, a simpler description is available.   The Hilbert space bundle $\V\to G_\M$ is homogeneous under the action of $G_L\times G_R$
on the base space $G_\M$, and we can pick either a $G_L$-invariant or a $G_R$-invariant trivialization of this  bundle.  Picking a $G_L$-invariant
trivialization can be accomplished by picking any trivialization at all over a chosen point $1\in G_\M$ and then extending this, by acting with $G_L$,
to a $G_L$-invariant trivialization over all of $G_\M$.  The extension exists and is unique because for any $g\in G_\M$, there is a unique element of $G_L$,
namely $g$, that maps $1\in G_\M$ to $g\in G_\M$.    This $G_L$-invariant trivialization is not $G_R$-invariant.   However, following the same procedure
with $G_R$ instead of $G_L$, we can define a $G_R$-invariant trivialization that is not $G_L$-invariant.

Either way, once we trivialize the bundle in a $G_L$- or $G_R$-invariant fashion, there is a simple description of the Hilbert space $\h\H_\TFD$ and of the
action of $G_L\times G_R$ on it.   When the bundle is trivialized, $\h\H_\TFD$ becomes a simple tensor product $\h\H_\TFD=\H_\TFD\otimes L^2(G_\M)$.
Thus a wavefunction becomes an $\H_\TFD$-valued function on $G_\M$.    However, there are two such  formalisms, depending on
whether we use the left-invariant or right-invariant trivialization of $\V$.  Let us write $\Psi(g)$ for the wavefunction with the left-invariant trivialization
and $\chi(g)$ for the wavefunction with the right-invariant trivialization.   To describe concretely the action of $G_L\times G_R$ on $\Psi(g)$ or $\chi(g)$,
we will need to remember  the action of the diagonal group $G$ on $\H_\TFD$, generated by the operators $\h Q^a$ that were introduced earlier
and with the representation matrices $W(g)$.   

Let $S_{g_L,g_R}$ be the operator that represents a group element $g_L\times g_R\in G_L\times G_R$ if we use the $G_L$-invariant trivialization with wavefunction
$\Psi(g)$.   Concretely, $S_{g_L,g_R}(\Psi)$ is the function on $G_\M$ defined by
\be\label{tofo} (S_{g_L,g_R}(\Psi))(g)= W(g_R) \Psi(g_L^{-1} g g_R). \ee
The idea of this formula is that since $\Psi$ is defined with a $G_L$-invariant trivialization of the fibration, $G_L$ acts only on the base space $G_\M$ of the fibration
 and not on the fiber $\H_\TFD$, but as the trivialization is not $G_R$-invariant, $G_R$ acts on the fiber (via the operator $W(g_R)$) as well as on the base.
 The reader can verify that eqn. (\ref{tofo}) does define an action of $G_L\times G_R$ on $\h\H_\TFD$.   If instead we use a wavefunction $\chi(g)$ defined
 with the $G_R$-invariant trivialization, then the roles of $G_L$ and $G_R$ are reversed.    The action of $G_L\times G_R$ is then described by operators
 $T_{g_L,g_R}$ such that
 \be\label{ofo} (T_{g_L,g_R}(\chi))(g) = W(g_L) \chi(g_L^{-1} g g_R). \ee
 The relation between the two formalisms is simply $\chi(g)=W(g)\Psi(g)$, or 
 \be\label{mofo}\chi=\Lambda\Psi, \ee
 where $\Lambda:\h\H_\TFD\to \h\H_\TFD$ is the defined by $(\Lambda\Psi)(g)=W(g)\Psi(g)$.
 The equivalence of the two formalisms is defined by
 \be\label{loopy} T_{g_L,g_R}=\Lambda S_{g_L,g_R}\Lambda^{-1}. \ee
 
 Let $t_L^a$ and $t_R^a$ be the generators of the $G_L$ and $G_R$ actions on $L^2(G_\M)$.   Thus in view of the formulas of the
 last paragraph,  in the formalism based on a $G_L$-invariant trivialization, 
 the action of $G_L$ on $\h\H_\TFD=\H_\TFD\otimes L^2(\R)$ is generated by $t^a_L$, while the action of $G_R$ is generated by $t_R^a+Q^a$.
In the formalism based on a $G_R$-invariant trivialization, the generators are instead $t_L^a+Q^a$ and $t_R^a$, respectively.
 
Now we can explain how to incorporate the collective coordinates associated with the $G$ symmetry in the algebra of observables outside the black hole
horizon.   On the right side of the horizon, for a fixed choice of the collective coordinates, we have an algebra $\A_{r,0}$ acting on $\H_\TFD$.   To incorporate
the collective coordinates in the left-invariant formalism, 
we replace $\H_\TFD$ by $\h\H_\TFD=\H_\TFD\otimes L^2(G_\M)$, and we adjoin the $G_R$ generators $t_R^a+Q^a$ to $\A_{r,0}$.
In other words, $\A_R$ is the algebra generated by $\A_{r,0}$ together with $t_R^a+Q^a$.   What we have just arrived at is the mathematical definition of the
crossed product of the algebra $\A_{r,0}$ by a group of automorphisms $G$.   We denote it as $\A_R=\A_{r,0}\rtimes G$.

As for $\A_L$, it is the commutant of $\A_R$.   It is generated by $t_L^a$ and $\Lambda^{-1} \A_{\ell,0}\Lambda$.   

This description treats the left and right sides of the black hole asymmetrically because of using a trivialization that is $G_L$-invariant but not $G_R$-invariant.
We can reverse the roles of $G_L$ and $G_R$ by conjugating by $\Lambda$.   Then we get a description in which $\A_L$ is generated by $\A_{\ell,0}$ and
$t_L^a+Q^a$, while $\A_R$ is generated by $t_R^a$ and $\Lambda \A_{r,0}\Lambda^{-1}$.      In all cases, regardless of the formalism, $\A_L$ and $\A_R$
are isomorphic to the crossed product of $\A_{\ell,0}$ or $\A_{r,0}$ by $G$.
 
 When we got to essentially this point in the discussion of time translations in section \ref{beyond}, we observed that there exists a left-right symmetric
 formalism obtained by conjugating by $\Lambda^{1/2}$ (eqn. (\ref{elb})). In the case of a nonabelian group $G$, there is  no equally convenient
 analog of $\Lambda^{1/2}$.
 
 Thus collective coordinates associated to an automorphism group $G$ 
 can always be included in the algebra of observables to the right (or left) of the eternal black hole horizon by replacing
 the  algebra of observables $\A_{r,0}$  defined in a particular background with its crossed product $\A_R=\A_{r,0}\rtimes G$.   However, in contrast
 to the case of time translation symmetry, it seems that the crossed product with a compact automorphism group does not lead to a qualitative change in the
 algebraic structure.  In quantum field theory, for compact $G$, $\A_{r,0}\rtimes G$ is always an algebra of Type III$_1$, like $\A_{r,0}$.  A proof of this
 has been sketched by R. Longo.   The main input is the fact that $\H_{\TFD}$ contains all representations of $G$, each with infinite multiplicity.

\vskip1cm
 \noindent {\it {Acknowledgements}}  I am greatly indebted to R. Longo for many patient explanations  and in particular for explaining
 the properties of the crossed product.   I also
 thank M. R. Douglas, H. Liu, and R. Mahajan for helpful discussions, N. Lashkari for assistance with the Connes cocycle, and V. Chandrasekharan
 for a careful reading of the manuscript and pointing out some inaccuracies.
 Research supported in part by NSF Grant PHY-1911298.
 \bibliographystyle{unsrt}

\end{document}